\documentstyle[12pt,epsf]{article}

\catcode `\@=11 \@addtoreset{equation}{section}

\newcommand{\plot}[1]
{\begin{center}
\epsfysize=6cm 
\vspace{-5mm}
\parbox{\epsfxsize}{\epsffile{#1}}
\vspace{5mm}
\end{center}}

\def\be{\begin{equation}}
\def\ee{\end{equation}}
\def\l{\label}
\def\bohrs{\mbox{ bohrs}}

\def\eV{\mbox{ eV}}

\def\au{\mbox{ a.u.}}

\def\A{{\cal A}}
\def\C{{\cal C}}
\def\E{{\cal E}}
\def\S{{\cal S}}

\def\rbt{{r_{b2}}}
\def\rbo{{r_{b1}}}
\def\rao{{r_{a1}}}

\def\rc{{r_{c}}}
\def\r{{r_{12}}}
\def\g{\gamma}
\def\k{\kappa}
\def\tone{\theta_1}
\def\po{\varphi_1}
\def\tt{\theta_2}
\def\pt{\varphi_2}
\def\erf{{\mbox{erf}}}
\def\erfc{{\mbox{erfc}}}
\begin{document}
\begin{titlepage}
\title{\Large\bf ON VARIATIONAL SOLUTION OF THE FOUR-BODY 
SANTILLI-SHILLADY MODEL OF $H_2$ MOLECULE}
\author{{\normalsize\bf A.K.Aringazin}\\
{\normalsize Karaganda State University, Karaganda 470074 Kazakstan}\\
{\normalsize ascar@ibr.kargu.krg.kz}\\
}
\date{December 1999}
\maketitle
\abstract{
In this paper, we apply Ritz variational approach to a new
isochemical model of $H_2$ molecule suggested by Santilli and Shillady.
We studied Gaussian, $V_g$, and exponential, $V_e$, screened
Coloumb potential {\it approximations}, as well as the original 
Hulten potential, $V_h$, case. 
Both the Coloumb and exchange integrals have been calculated only for $V_e$
owing to Gegenbauer expansion while for $V_g$ and $V_h$
cases we achieved analytical results only for the Coloumb integrals.
We conclude that the $V_e$-based model is capable 
to fit experimental data on $H_2$ molecule in confirmation of the 
results of numerical HFR approach by Santilli and Shillady.
Also, we achieved the energy-based estimation of the weight 
of the isoelectronium
phase which is appeared to be of the order of 1\%...6\%, 
for the case of $V_e$. However, we note that this is {\it not} 
the result corresponding to the original Santilli-Shillady model,
which is based on the Hulten potential $V_h$.
An interesting result is that in order to prevent
divergency of the Coloumb integral for $V_h$
the correlation length parameter $r_c$ should run discrete set of values.
We used this condition in our $V_e$-based model. 
}
\end{titlepage}

\section{Introduction}

In this paper, we consider the four-body Santilli-Shillady isochemical 
model of $H_2$ molecule \cite{SS, SS2, Santilli} characterized by 
additional short-range attractive Hulten potential between the electrons.
This potential is assumed to lead to bound state of electrons
called isoelectronium.
The restricted three-body Santilli-Shillady model 
(stable and point-like isoelectronium) of $H_2$ has been studied in 
ref. \cite{ibr6}, in terms of exact solution. For the mass of isoelectronium
$M=2m_e$, this solution implied much lower energy
than the experimental one so we varied the mass and obtained 
that $M=0.308381m_e$ fits the experimental binding energy, 
$E_{exper}[H_2]=-1.174474 \au$ up to six decimal places, 
although at bigger value of the internuclear 
distance, $R=1.675828 \au$ in contrast to $R_{exper}[H_2]=1.4011 \au$
We realize that the three-body model is capable to represent 
the binding energy but it is only some approximation to the four-body model, 
and one should study the general four-body hamiltonian of the 
Santilli-Shillady model as well.

In the present paper, we use Ritz variational approach to 
the four-body Santilli-Shillady isochemical model of $H_2$ molecule,
i.e. without restriction that the isoelectronium is stable 
and point-like particle, in order to find the ground state energy 
and bond length of the $H_2$ molecule. 

In Sec.~2, we analyze some features of the four-body Santilli-Shillady
isochemical model of $H_2$ molecule.

In Sec.~3, we apply Ritz variational approach to the four-body 
Santilli-Shillady model of $H_2$ molecule.
We calculate Coloumb integral for the cases of Hulten potential 
(Sec.~3.1.1), exponential screened Coloumb potential (Sec.~3.1.2), 
and Gaussian screened Coloumb potential (Sec.~3.1.3). 
Owing to Gegenbauer expansion, exchange integral has 
been calculated for the case of exponential screened potential, 
with some approximation made (Sec.~3.1.4).
Exchange integrals for the Hulten potential and the Gaussian screened 
Coloumb potential have not been derived, and require more study.
We present main details of calculations of the Coloumb and exchange
integrals which have been appeared to be rather 
cumbersome, especially in the case of Hulten potential. 

In Sec.~3.2, we make numerical fitting of the variational energy 
for the case of exponential screened Coloumb potential $V_e$.  
Also, we estimate the weight of the isoelectronium phase.
However, we use all the important results of the analysis made for 
the Hulten potential $V_h$.

1) We conclude that the $V_e$-based model with the one-level 
isoelectronium is {\it capable to fit the experimental data on 
$H_2$ molecule} (both the binding energy $E$ and the bond length $R$).
This is in confirmation of the results of numerical HFR approach 
(SASLOBE routine) to the $V_g$-based model of ref. \cite{SS}.

2) One of the interesting implications of the Ritz variational approach 
to the Hulten potential case is that the correlation length parameter 
$r_c$, entering the Hulten potential, and, as a consequence,
the variational energy, should run discrete set of values
during the variation. In other words, this means that {\it only some fixed 
values of the effective radius of the one-level isoelectronium are admitted},
in the original Santilli-Shillady model, within the framework of 
the Ritz approach. This highly remarkable property 
is specific to the Hulten potential $V_h$ while it is absent in the $V_e$, or 
$V_g$-based models.

3) Also, we achieved an estimation of the {\it weight of the isoelectronium
phase} for the case of $V_e$-based model which is appeared to be of 
the order of 1\%...6\%. This weight
has been estimated from the {\it energy} contribution, related to the 
exponential screened potential $V_e$, in comparison to the contribution
related to the Coloumb potential. 

4) Another general conclusion is that the effective radius of the 
isoelectronium $r_c$ should be less that $0.25 \au$

We note that the weight of the phase does not mean directly a {time share} 
between the two regimes, i.e., 1...6\% of time for the pure isoelectronium 
regime, and 99...94\% of time for the decoupled electrons regime. 
This means instead {\it relative contribution to the total energy}
provided by the potential $V_e$ and by the usual Coloumb potential between 
the electrons, respectively.
As a consequence, the weight of the isoelectronium phase, which can be 
thought of as a measure of stability of the isoelectronium,  may be 
\begin{enumerate}
\item Different from the obtained 1...6\% when calculated
for some other characteristics of the molecule, e.g., for a relative 
contribution of the pure isoelectronium to the total magnetic moment of 
the $H_2$ molecule;
\item Different from the obtained 1...6\% for the case 
of the original Hulten potential $V_h$.
\end{enumerate}
So, the result of the calculation made in this paper is {not the final 
result} implied by the general four-body Santilli-Shillady model 
of $H_2$ molecule since the latter model is based on the Hulten 
potential $V_h$. 
This paper can be viewed only as a preliminary study to it.
However, we have made some essential advance in 
analyzing the original Hulten potential case (Sec.~3.1.1), which
we have used in the $V_e$-based model.

Below, we describe the procedure used in Sec.~3 in a more detail.
In Ritz variational approach, the main problem is to calculate
analytically so called molecular integrals. The variational molecular 
energy, in which we are interested in, is expressed in terms of 
these integrals; see Eq.(\ref{Emol}). These integrals
arise when using some wave function basis (usually it is a simple 
hydrogen ground state wave functions) in the Schr\"odinger equation
for the molecule. The main idea of the Ritz approach is to introduce
parameters into the wave function, and vary them, together with
the internuclear distance parameter $R$, to achieve a minimum of the 
molecular energy. 
In the case under study, we have two parameters,
$\gamma$ and $\rho$, where $\gamma$ enters hydrogen-like ground state
wave function (\ref{wf0}), 
and $\rho=\gamma R$ measures internuclear distance.
These parameters should be varied (analytically or numerically) 
in the final analytical expression of the molecular energy, 
after the calculation is made for the associated molecular integrals.

However, the four-body Santilli-Shillady model of $H_2$ molecule 
suggests additional, 
Hulten potential interaction between the electrons. The Hulten potential
contains two parameters, $V_0$ and $r_c$, where $V_0$ is a general
factor, and $r_c$ is a correlation length parameter which can be 
viewed as an effective radius of the isoelectronium; see Eq. (\ref{V_h}).
Thus, we have four parameters to be varied, 
$\gamma$, $\rho$, $V_0$, and $r_c$. 
The introducing of Hulten potential 
leads to modification of some molecular integrals, namely, of
the Coloumb and exchange integrals; 
see Eqs. (\ref{Coloumb2}) and (\ref{exchange2}). The other
molecular integrals remain the same as in the case of usual model
of $H_2$, and we use the known analytical results for them.
So, we should calculate the 
associated Coloumb and exchange integrals for the Hulten potential 
to get the variational energy analytically. 
In fact, calculating of these integrals,
which are six-fold ones, constitutes the main problem here. 
Normally, Coloumb integral, which can be performed in bispherical 
coordinates, is much easier than the exchange one, which is performed in 
bishperoidal coordinates.

Calculation of the Coloumb integral for Hulten potential, $V_h$, 
appeared to be rather nontrivial (Sec.~3.1.1). Namely, we used bispherical
coordinates, and have faced several special functions, such as 
polylogarithmic function, Riemann $\zeta$-function, digamma function, 
and Lerch function, during the calculation. Despite the fact that
we see no essential obstacles to calculate this six-fold integral, 
we stopped the calculation after fifth step because sixth (the last) step
assumes necessity to calculate it separately for each integer value of 
$\lambda^{-1} \equiv (2\gamma r_c)^{-1}$, together with the need to 
handle very big number of terms. During the calculations, we were
forced to use the condition that 
$\lambda^{-1}$ should take integer values in order to prevent
divergency of the Coloumb integral for Hulten potential. 
Namely, some combination of terms
containing Lerch functions gives a finite value only if this condition
holds. This condition is specific to Hulten potential. 
Note also that we can not get general form of a final expression 
for the Coloumb 
integral for Hulten potential because Lerch functions entering the 
intermediate expression (after the fifth step, see Eq.(\ref{Icalc})) 
can be integrated over only for a concrete numerical value of their 
third argument.

In order to proceed with the Santilli-Shillady approach, we 
invoke to two different simplified potentials, the exponential screened
Coloumb potential, $V_e$, and the Gaussian screened Coloumb potential,
$V_g$, instead of the Hulten potential $V_h$. 
They both mimic well Hulten potential
at short and long range asymptotics, and each contains two parameters,
for which we use the notation, $A$ and $r_c$. In order to 
reproduce the short range asymptotics of Hulten potential
the parameter $A$ should have the value $A=V_0r_c$, for both the potentials.
The Coloumb integrals for these two potentials have been calculated
{\it exactly} (Secs.~3.1.2 and 3.1.3) owing to the fact that they are 
much simpler than the Hulten potential. Particularly, we note that 
the final expression of the Coloumb integral for 
$V_g$ contains only one special function, the error function $\erf(z)$, 
while for $V_e$ it contains no special functions at all.

Having these results we turned next to the most hard
part of work: the exchange integral. Usually, to calculate it one
has to use bispheroidal coordinates, and needs in an expansion of the 
potential in some orthogonal polynomials, such as Legendre polynomials,
in bispheroidal coordinates. 
Here, only the exponential screened potential $V_e$ is known to have
such an expansion but it is formulated, however, in terms of bispherical
coordinates (the Gegenbauer expansion). 
Accordingly, we calculated exactly the exchange integral for $V_e$, 
at {\it zero} internuclear separation, $R=0$, at which case one can use
bispherical coordinates. After that, we recovered 
partially the $R$ dependence using the standard result for the 
exchange integral for Coloumb potential (Sugiura's result).
Thus, we achieved some approximate expression of the exchange 
integral for the case of $V_e$. So, all the subsequent results 
correspond to the $V_e$-based model.

Inserting obtained $V_e$-based Coloumb and exchange integrals into 
the total molecular energy expression, we get the final analytical expression
containing four parameters, $\gamma$, $\rho$, $A$, and $r_c$. 
Prior to going into details of the energy minimization for the 
$V_e$-based (approximate) model, we
analyze the set of parameters, and the conditions which we derived in
the original Hulten potential case. 

(1) From the analysis of Hulten potential, we see (Sec.~2.1) that the 
existence of a bound state of two electrons (which is proper isoelectronium) 
leads to the following relationship between the parameters
for the case of {\it one} energy level of the electron-electron system: 
$V_0= \hbar^2/(2m r_c^2)$. So, using the above mentioned 
relation $A=V_0r_c$ we have $A=1/r_c \equiv 2\gamma/\lambda$, 
in atomic units ($\hbar=m_e=c=1$). Thus note that, in this paper,
we confined our consideration to the case of {\it one-level} 
isoelectronium.

(2) From the analysis of the Coloumb integral for Hulten potential,
we see (Sec.~3.1.1) that the condition, 
$\lambda^{-1}=$ {\it integer number}, should hold, and one can use it
as well.

We use the above two conditions, coming from the Hulten potential
analysis, in the energy minimization calculations 
for the case of our $V_e$-based model.
The first condition diminishes the number of independent parameters
by one (they become three, $\gamma$, $\rho$, and $\lambda$)
while the second condition means a discretization of the $\lambda$
parameter, $\lambda^{-1}= 4,5,6,\dots$  Here, we used the condition
$\lambda^{-1}>3$ which we obtained during the calculation of
the Coloumb integral for $V_e$.

With the above set up, we minimized the total molecular energy 
of the $V_e$-based model. Numerical
analysis shows that the $\lambda$ dependence does not reveal
any minimum, in the interval of interest, $4<\lambda^{-1}<60$, 
while we have a minimum of the energy at some 
values of $\gamma$ and $\rho$. So, we calculated the energy
minima for different values of $\lambda$, in the interval of interest,
$4<\lambda^{-1}<60$. Results are presented in Tables~2 and 3.
One can see that the binding energy decreases with the increase of 
the parameter $r_c$, which corresponds to the effective radius of the 
isoelectronium.

The following remarks are in order.

(i) Note that the discrete character of $r_c$ does not mean that 
the isoelectronium is some kind of a multilevel system, with different 
effective radia of isoelectronium assigned to the levels. 
We remind that we treat the
isoelectronium as {\it one-level} system due to the above mentioned
relation $V_0=\hbar^2/(2m r_c^2)$. In fact, this means that there 
is a set of one-level isoelectronia of different fixed effective radia
from which we should select only one, to fit the experimental data.

(ii) The use of the exponential screened potential $V_e$ can only be 
treated as some {\it approximation} to the original Hulten potential,
and, thus, to the original Santilli-Shillady model of $H_2$ molecule.
So, the numerical results obtained in Sec.~3.2 are valid only 
within this approximation. Hulten potential makes a difference
(one can see this, e.g., by comparing Sec.~3.1.1 and Sec.~3.1.2), and  
it is worth to be investigated more closely by, for example, 
combination of analytical and numerical methods.

(iii) The results obtained in ref. \cite{SS} are based on the Gaussian
screened Coloumb potential $V_g$ approximation, to which the present 
work gives support in the form of exact analytical calculation of 
the Coloumb  integral for $V_g$ (Sec.~3.1.3). Also, the present work 
gives possibility to make a comparative analysis of ref. \cite{SS}, 
due to some similarity of the used potentials, $V_e$ and $V_g$.

(iv) Both the Coloumb integrals, for $V_e$ and $V_g$, reveal a minimum  
in respect with $\lambda=2\gamma r_c$, i.e. in respect with $r_c$ 
(see Figures~6 and 9) since minimization in Ritz parameter $\gamma$ 
is made independently. 
In principle, this gives us an opportunity to minimize the total molecular 
energy $E_{mol}$ with respect to $r_c$. However, there are two reasons that 
we can not provide this minimization. First, these minima correspond 
to rather large values of $r_c$, namely, $r_c \geq 1 \au$ for $V_e$ (Fig.~6), 
and $r_c>2 \au$ for $V_g$ (Fig.~9).
Of course, this is not an obstacle to do minimization but we note that we 
generally assume that the effective radius of the isoelectronium $r_c$ is 
much less than the internuclear distance, 
$r_c \ll R=R_{exper}[H_2]=1.4011 \au$ 
Second, and the main, reason is that for the exponential screened potential 
case (Sec.~3.1.2) the parameter $\lambda$ should be less than 1/3 to provide 
convergency of the associated Coloumb integral. Typically, 
$\gamma \simeq 1.2$, from which we obtain the condition 
$r_c=\lambda/2\gamma < 0.2 \au$
Also, for the Hulten potential case (Sec.~3.1.1), we obtained 
$\lambda<1/2$, and hence $r_c < 0.25 \au$ 
This means that, in fact, it is {\it impossible} to reach finite minimum of 
the total molecular energy $E_{mol}$ in respect with $r_c$ since the Coloumb 
integrals blow up, at $r_c>0.25 \au$, leading thus to infinite total energy 
$E_{mol}$. So, in our approach we arrive at a strict theoretical conclusion 
that the effective radius of the isoelectronium $r_c$ must be less than 
$0.25 \au$ Clearly, this supports our assumption that $r_c$ is much less than
the internuclear distance $R$.

\section{Santilli-Shillady model and the barrier}\l{H-C}
\setcounter{equation}{0}

In this Section, we consider the general four-body 
Santilli-Shillady model \cite{SS} of $H_2$ molecule, 
in Born-Oppenheimer approximation (i.e. at fixed nuclei).
Shr\"odinger equation for $H_2$ molecule with the additional 
short range attractive Hulten potential between the electrons
is of the following form:
\begin{eqnarray}\l{general2}
\left( -\frac{\hbar^2}{2m_1}\nabla^2_1
-\frac{\hbar^2}{2m_2}\nabla^2_2
-V_0\frac{e^{-r_{12}/r_c}}{1-e^{-r_{12}/r_c}} +\frac{e^2}{r_{12}} 
\right.
\\ \nonumber
\left.
-\frac{e^2}{r_{1a}}  -\frac{e^2}{r_{2a}}
-\frac{e^2}{r_{1b}}  -\frac{e^2}{r_{2b}}
+\frac{e^2}{R}
\right)|\psi\rangle = E|\psi\rangle,
\end{eqnarray}
where $R$ is distance between the nuclei $a$ and $b$.

Interaction between the two electrons in the model
is due to the potential
\be\l{V12}
V(r_{12}) = V_C(r_{12})+V_h(r_{12})
=\frac{e^2}{r_{12}}-V_0\frac{e^{-r_{12}/r_c}}{1-e^{-r_{12}/r_c}},
\ee
where $r_{12}$ is distance between the electrons,
$V_0$ and $r_c$ are real positive parameters.
Here, first term, $V_C$, is usual repulsive Coloumb potential,
and the second term, $V_h$, is an attractive Hulten potential.

Extrema of $V(r_{12})$ are defined by the equation
\be\l{extremum}
V'(r_{12}) = -\frac{e^2}{r^2_{12}} 
+\frac{V_0}{r_c}\frac{e^{r_{12}/r_c}}{(e^{r_{12}/r_c}-1)^2} =0.
\ee
In the limit $r_{12}\to\infty$, potential 
$V(r_{12})\sim e^2/r_{12}=V_C(r_{12})$.
Series expansion of $V(r_{12})$ at $r_{12}\to 0$ is 
\be\l{series}
V(r_{12})|_{r_{12}\to 0}=\frac{e^2-V_0r_c}{r_{12}} 
+\frac{V_0}{2}-\frac{V_0}{12r_c}r_{12} +O(r^3_{12}).
\ee
In general, there is relationship of Hulten potential to
Bernoulli polynomials $B_n(x)$. Namely, Bernoulli polynomials 
are defined due to 
\be\l{Bernoulli}
\frac{se^{xs}}{e^s-1}=\sum\limits_{n=0}^{\infty} B_n(x)\frac{s^n}{n!},
\ee
and we can reproduce Hulten potential,
\be\l{HultenBernoulli} 
\frac{e^{s}}{1-e^s}
=-\frac{1}{s}\sum\limits_{n=0}^{\infty} B_n(1)\frac{s^n}{n!},
\ee
taking $s=-r_{12}/r_c$. First five Bernoulli coefficients are
\be
B_0(1)=1,\  B_1(1)=\frac{1}{2},\  B_2(1)=\frac{1}{6}, \ 
B_3(1)=0,\  B_4(1)=-\frac{1}{30}.
\ee
Eq.(\ref{HultenBernoulli}) means expansion of Hulten potential
with the use of Bernoulli coefficients.

Eq.(\ref{series}) implies that to have an {\it attraction} near $r_{12}=0$,
which is necessary for forming of isoelectronium,
we should put the condition
\be\l{condition1V0rc}
V_0r_c > e^2.
\ee
We note that, in view of the asymptotics (\ref{series}), 
$Q=\sqrt{V_0r_c}$ can be thought of as {\it Hulten charge} of the electrons.

Under this condition, $V(r_{12})$ has one maximum at the point
defined by Eq.(\ref{extremum}). 
This is the equilibrium point at which
the Coloumb potential is equal to the Hulten potential.
So, we have barrier ($B$) separating two asymptotic regions,
($A$) $r\to 0$ and ($C$) $r\to \infty$, with 
Coloumb-like attraction and Coloumb-like repulsion, 
respectively.

In the region $A$, attractive Hulten potential $V_h$ dominates,
and therefore two electrons form bound state
(isoelectronium),
while in the region $C$ Coloumb repulsion $V_C$ dominates,
and they are separated. This separation is limited
by the size of the neutral molecule. For example, assuming
that $H_2$ molecule is in the ground state we have
$r \leq r_{mol} = 3.46 \bohrs$, where we have assumed that 
separation between the protons is $R=1.46 \bohrs=0.77 \AA$.

Existence of the bound state of the electrons and 
of the barrier $B$ is a novel feature
provided by the model. The asymptotic states,
in regions $A$ and $C$, pertube each other
due to the barrier effect in region $B$.

\subsection{Region $A$}\l{SecA}

In the case 
\be\l{condition2V0rc}
V_0r_c \gg e^2
\ee
we can ignore Coloumb repulsion $V_C$, and region $A$ is a Hulten region,
$|V_h| \gg |V_C|$; see Eq.(\ref{series}).
Then, exact solution of one-particle 
Schr\"odinger equation with Hulten potential $V_h$,
where wave function has the boundary conditions
$\psi(0)=0$ and $\psi(\infty)=0$ (see \cite{Flugge}, problem 68),
can be used to establish 
the relation between the parameters $V_0$ and $r_c$,
and to estimate $r_c$. 

Energy spectrum for Hulten potential is given by
\be\l{energyHulten}
E_n = -V_0\left( \frac{\beta^2-n^2}{2n\beta}\right)^2,
\quad n=1,2,\dots.
\ee
where
\be\l{beta}
\beta^2 = \frac{2mV_0}{\hbar^2}r_c^2,
\ee
and $m$ is mass of the particle.
Assuming that there is only {\it one} energy level, namely,
ground state $n=1$, we obtain the condition
\be\l{conditionbeta}
\beta^2=1,
\ee
which can be rewritten as
\be\l{r-c}
r_c = \hbar\sqrt{\frac{1}{2mV_0}}.
\ee
Note that this state is characterized by approximately
{\it zero} energy, $E_1=0$, due to Eq.(\ref{energyHulten});
strictly speaking, 
$\beta^2$ must be bigger but close to 1 in Eq.(\ref{conditionbeta}). 

We should to note that the number of energy levels for 
Hulten potential is always finite due to Eq.(\ref{energyHulten}). 
Assumption that there are more than one energy levels
in the bound state of two electrons, i.e. that 
$\beta>1$, leads to drastical decrease of ground level energy
$E_1<0$, and relatively small increase of characteristic size 
of isoelectronium in the ground state.

As the conclusion, the model implies "quantization" of 
the distance between two electrons, $r=r_{12}$, 
namely, forming of relatively small quasiparticle
(isoelectronium) characterized by total mass $M=2m_e$,
charge $q=-2e$, spin zero, $s=0$,  and small size 
in the ground one-level state.
This quasiparticle, as a strongly correlated system of 
two electrons, moves in the potential of two protons
of $H_2$ molecule, and one can apply methods developed for
$H_2^+$ ion, with electron replaced by isoelectronium, to calculate 
approximate energy spectrum of $H_2$ \cite{ibr6}.
However, this quasiparticle is not stable, being a quasi-stationary 
state, due to finite height and width of the barrier $B$. So, we must take 
into account effects of both regions $B$ and $C$ to obtain correct
energy spectrum of $H_2$ molecule, within the framework
of the model.

\subsection{Region $B$}



Quasiclassically, due to smooth shape of the barrier,
and because of exponential decrease of wave functions 
inside the barrier, electrons are not much time in region $B$, 
so we can ignore this {\it transient phase} in subsequent 
consideration.

We should to point out that the existence of the bound state
in the region $A$ and repulsion in the region $C$ 
unavoidably leads to existence of the barrier.

\subsection{Region $C$}

In general, region $C$ is infinite, $r_{C}<r<\infty$,
where $r_C$ is the distance between two electrons
at which the Hulten potential is much smaller than
the Coloumb potential, $|V_h|\ll|V_C|$. 

In this region, electrons are not strongly correlated,
in comparison to that in region $A$. Here, correlation
is due to usual overlapping, Coloumb repulsion, exchange effects,
and Coloumb attraction to protons.
Shortly, we have the usual set up as it for the standard
model of $H_2$ molecule.

Discarding, for a moment, effects coming from the consideration
of regions $A$ and $B$, we have finite motion of the
electrons in region $C$. Namely, in the ground state of $H_2$, 
the distance between electrons is confined by 
$r = r_{mol}=3.46 \bohrs$.
We restrict consideration by the ground state of $H_2$ molecule.

Due to this finiteness of the region $C$, $r<r_{mol}$,
two electrons on the same orbit have constant probability 
to penetrate the barrier to form strongly correlated system,
isoelectronium, and vice verca.

\subsection{Model of decay of isoelectronium}

Below, we assume that the isoelectronium undergoes decay, and the 
resulting two electrons are separated by sufficiently large distance,
in the final state. This leads us to consideration of
the {\it effective life-time of isoelectronium}.
To estimate the order of the life-time,
we use ordinary formula for radioactive $\alpha$-decay 
since the potential $V(r)$ is of the same shape, with 
very sharp decrease at $r<r_{max}$ and Coloumb repulsion at
$r>r_{max}$. This quisiclassical model is a crude approximation 
because in fact the electrons do not leave the molecule.
Moreover, we have the two asymptotic regimes simultaneously,
with some distribution of probability,
and it would be more justified here to say on frequency of the 
decay-formation process. 
However, due to our assumption of small size of isoelectronium,
in comparison to the molecule size, we can study an elementary
process of decay separately, and use the notion of life-time.

Decay constant is
\be
\lambda =\frac{\hbar D_0}{2mr_{max}^2}
\exp\left\{ 
-\frac{4\pi Z e^2}{\hbar}\sqrt{\frac{m}{2E}}
+\frac{4e}{\hbar}\sqrt{Zmr_{max}}
\right\},
\ee
where we put, in atomic units, 
\be
\hbar=1, \ e=1, \ m=1/2, \  Z=1, \ r_{max}=0.048, \ E=1.
\ee
Here, $E=1 \au = 27.212 \eV$ is double kinetic energy of the electron 
on first Bohr's orbit, $a_0=0.529 \AA$, 
that corresponds approximately to maximal relative kinetic 
energy of two electrons in ground state of $H_2$, and $m=1/2$
is reduced mass of two electrons.

We obtain the following numerical estimation for the life-time
of isoelectronium:
\be
1/\lambda = D_0\cdot1.6\cdot10^{-17} \sec,
\ee
i.e. it is of the order of 1 atomic unit of time,
$\tau = 2.42\cdot10^{-17} \sec$. For lower values of the 
relative energy $E$, we obtain longer lifetimes; see Table~2.

The quasiclassical model for decay we are using here is the following.
Particle of reduced mass $m=1/2$ penetrate the barrier $B$.
This means a decay of isoelectronium. 
In the center of mass of electrons system,
electrons undergo Coloumb repulsion and move in opposite directions 
receiving equal speed so that at large distances, $r\gg r_{max}$,
each of them have some kinetic energy.
This energy can be given approximate upper estimation using linear velocity
of electron on first Bohr's orbit, $v=2.19\cdot 10^{6}$ cm/sec,
since electrons are in the ground level of $H_2$ molecule
(this is the effect of the nuclei).
This upper estimation corresponds to assumption of zero velocity of 
the center of mass in respect to protons which we adopt here.
Kinetic energy of the particle of reduced mass is then
double kinetic energy of electron, in center of mass system.

\begin{table}[thp]\l{TableLifetime}
\begin{center}
\begin{tabular}{|c|c|c|}
\hline
Energy $E$, a.u. & eV  & Lifetime, $D_0\cdot$sec  \\
\hline
2           & 54.4        & $2.6\cdot10^{-18}$\\
\hline
1           & 27.2        & $1.6\cdot10^{-17}$\\
\hline
0.5         & 13.6        & $2.2\cdot10^{-16}$\\
\hline
0.037       & 1           & $5.1\cdot10^{-6}$\\
\hline
0.018       & 0.5         & 4.0              \\
\hline
0.0018      & 0.1         & $3.1\cdot10^{+25}$\\
\hline
\end{tabular}
\caption{Lifetime of isoelectronium.
$E$ is relative kinetic energy of the electrons,
at large distances, $r\gg r_{max}$, in the center of mass system.}
\end{center}
\end{table}

%

As the conclusion, in the framework of the model,
$H_2$ molecule can be viewed as a mixed state of 
$H_2^+$ ion like system, i.e. {\it strongly correlated phase}
(Hulten phase), 
when electrons form isoelectronium,
and standard model of $H_2$, i.e. {\it weakly correlated phase}
(Coloumb phase),
when electrons are separated by large distance, $r>r_{max}$.
Note that, as it has been mentioned above,
we ignore the {\it transient phase} (inside the barrier)
in this consideration.
Evidently, the (statistical) weight of each phase depends on 
the characteristics of the potential $V(r_{12})$. 

For extremally high barrier, only one of the phases could be realized
with some energy spectra in each phase, namely, either spectrum of 
$H_2^+$ ion like system (with electron replaced by isoelectronium), 
or usual spectrum of $H_2$ molecule (without Hulten potential), 
respectively.

For high but finite barrier, each phase receives perturbation, 
and their (ground) energy levels split to two levels corresponding
to {\it simultaneous} realization of both the phases. Note that
the value $V_{max}$ is indeed high, $V_{max}\sim 500 \eV$,
under the given values of the parameters.

In general, existence of the strongly correlated phase
(isoelectronium) leads to {\it increase} of the predicted
dissociation energy, $D$, of $H_2$ molecule. Indeed, 
the mutual infuence of the regions $A$ and $C$ decreases
the ground energy level $E$ of $H_2$ due to the above mentioned
splitting. The general formula for $D$ is
\be
D=2E_0 -(E+\frac{1}{2}\hbar\omega),
\ee
where $2E_0=-1$ is total energy of two separated $H$ atoms,
and $\frac{1}{2}\hbar\omega$ is zero mode energy of the 
protons in $H_2$. So, decreasing of $E<0$ causes increase
of $D$.

It is remarkable to note that 
experimental data give dissociation energy $D_{exper}[H_2]=4.45 \eV$ 
for $H_2$ molecule (see, e.g. \cite{Flugge} and references therein) 
while theoretical predictions within the standard model are
$D=2.90 \eV$ (Heitler-London), $D=3.75 \eV$
(Flugge), and $D=4.37 \eV$ (Hylleraas). We observe that
improvement of the variational approximation gives better fits
but still it gives {\it lower} values (about 2\% lower) partially due to 
the fact that variational technique used there predicts 
generally bigger value (upper limit) for the ground energy.

Below, we use the same Ritz variational technique as it had been
used by Heitler, London and Hylleraas but the feature of the model is 
the existence of additional attractive short range potential between 
the electrons suggested by Santilli and Shillady.

\section{Variational solution for ground state energy of $H_2$ molecule}

In the limiting case of large distances between the nuclei,
$R\to \infty$, we have the total wave function of the electrons
in the form
\be\l{wf}
|\psi\rangle = f(r_{a1})f(r_{b2}) \pm f(r_{b1})f(r_{a2}),
\ee
where the first term corresponds to the case when 
electron 1 is placed close to nucleus $a$ and $f(r_{a1})$ 
is wave function of the corresponding separate $H$ atom while 
the second term corresponds to the case when electron 1 is placed 
close to nucleus $b$.
Symmetrized combination ($'+'$ sign) corresponds to antiparallel spins
of the electrons 1 and 2, and, as the result of the usual analysis,
leads to attraction between the $H$ atoms.
Below, we use this symmetrized representation of the 
total wave function as the approximation to exact wave function.

\subsection{Analytical calculations}

By using Ritz variational approach and representation (\ref{wf}), 
we obtain from the Schr\"odinger equation
(\ref{general2}) the energy of $H_2$ molecule in the following form 
(cf. \cite{Flugge}), 
\be\l{Emol}
E_{mol}
= 2\frac{\A+\A'\S}{1+\S^2} 
- \frac{2(\C+\E\S)-(\C'+\E')}{1+\S^2}
+\frac{1}{R},
\ee
where 
\be\l{overlap}
\S = \int dv\ f^*(r_{a1})f(r_{b1})
\ee
is overlap integral,
\be\l{Coloumb1}
\C = \int dv\ \frac{1}{r_{b1}}|f(r_{a1})|^2,
\ee
\be\l{Coloumb2}
\C' = \int dv_1dv_2\ 
\left(\frac{1}{r_{12}}-V_0\frac{e^{-r_{12}/r_c}}{1-e^{-r_{12}/r_c}}\right)
|f(r_{a1})|^2|f(r_{b2})|^2,
\ee
are Coloumb integrals,
\be\l{exchange1}
\E = \int dv\ \frac{1}{r_{a1}}f^*(r_{a1})f(r_{b1}),
\ee
\be\l{exchange2}
\E' = \int dv_1dv_2\ 
\left(\frac{1}{r_{12}}- V_0\frac{e^{-r_{12}/r_c}}{1-e^{-r_{12}/r_c}}\right)
f^*(r_{a1})f(r_{b1})f^*(r_{a2})f(r_{b2})
\ee
are exchange integrals, 
\be\l{A}
\A = \int dv\ 
f^*(r_{a1})\left(-\frac{1}{2}\nabla_1^2-\frac{1}{r_{a1}}\right)f(r_{a1})
\ee
and
\be\l{A'}
\A' = \int dv\ 
f^*(r_{a1})\left(-\frac{1}{2}\nabla_1^2-\frac{1}{r_{b1}}\right)f(r_{b1}).
\ee
We use atomic units, $e=1$, $m_1=m_2=m_e=1$.

Quite natural choice is that the wave functions in 
Eq.(\ref{wf}) are taken in the form of hydrogen ground state wave function,
\be\l{wf0}
f(r) = \sqrt{\frac{\g^3}{\pi}}e^{-\g r},
\ee
where $\g$ is Ritz variational parameter ($\g$=1 for the proper hydrogen
wave function), and $r=r_{a1}, r_{b1}, r_{a2}, r_{b2}$. 
With the help of $\g$
we should make better approximation to an exact wave function
of the ground state. Namely,
we should calculate all the integrals presented above analytically, 
and then vary the parameters $\g$ away from the value $\g=1$ 
and $R$ in some appropriate region, say $1<R<2$, 
to minimize the energy (\ref{Emol}). 
As the energy minimum will be identified the found 
value of the parameter $R$ corresponds to optimal distance
between the nuclei. This value should be compared to the experimental
value of $R$.

All the molecular integrals (\ref{overlap})-(\ref{A'}), 
except for the Hulten potential parts in (\ref{Coloumb2}) and 
(\ref{exchange2}), are wellknown and can be calculated exactly; 
see, e.g. \cite{Flugge}. Namely, they are
\be\l{OverlapS}
\S = \left(1+\rho+\frac{1}{3}\rho^2\right)e^{-\rho},
\ee
%
\begin{figure}
\plot{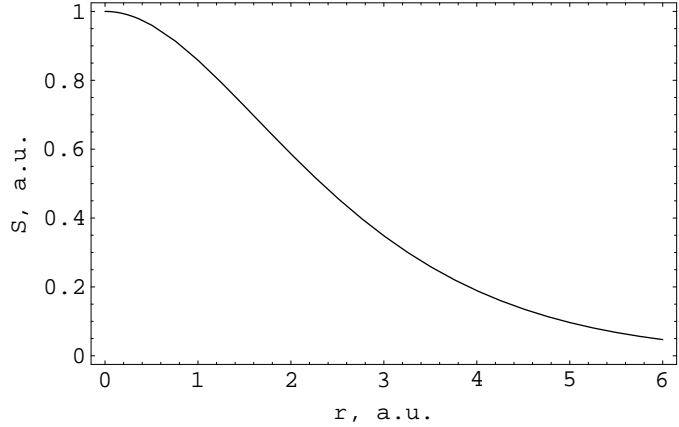}
\caption{The overlap integral $\S$ as a function of $\rho$, 
Eq. (\ref{OverlapS}). Here, $\rho=\gamma R$, where 
$\gamma$ is Ritz parameter and $R$ is the internuclear distance.}
\label{Fig1}
\end{figure}
\be\l{ColoumbCC}
\C \equiv \C_C= \frac{\g}{\rho}(1-(1+\rho)e^{-2\rho}),
\ee
%
\begin{figure}
\plot{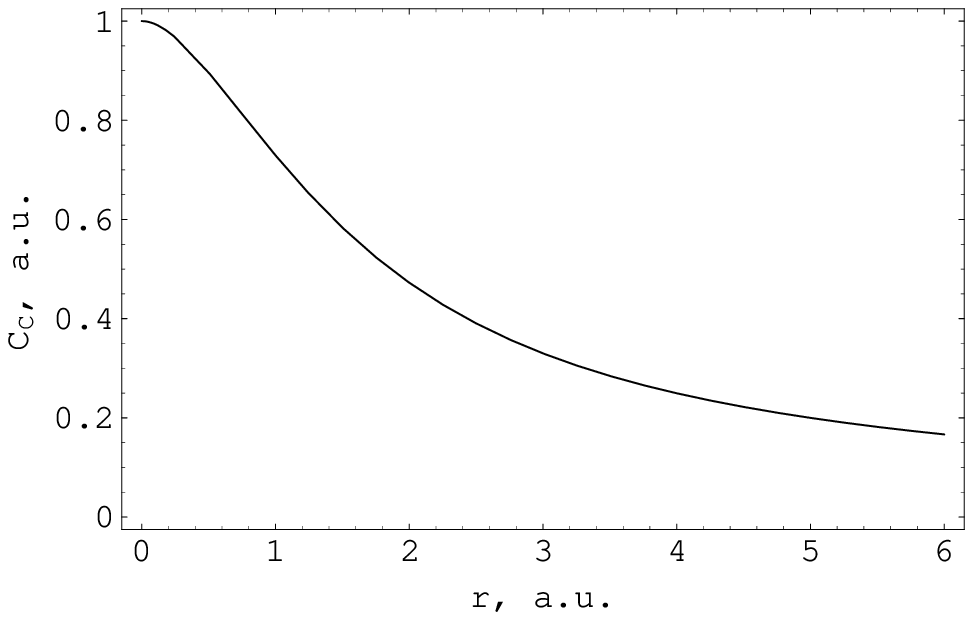}
\caption{The Coloumb integral $\C_C$ as a function of $\rho$, 
Eq. (\ref{ColoumbCC}).}
\label{Fig2}
\end{figure}
\be\l{C'0}
\C'_C \equiv \C'_{|V_0=0} = \frac{\g}{\rho}
\Bigl(
1-(1+\frac{11}{8}\rho+\frac{3}{4}\rho^2+\frac{1}{6}\rho^3)e^{-2\rho}
\Bigr),
\ee
%
\begin{figure}
\plot{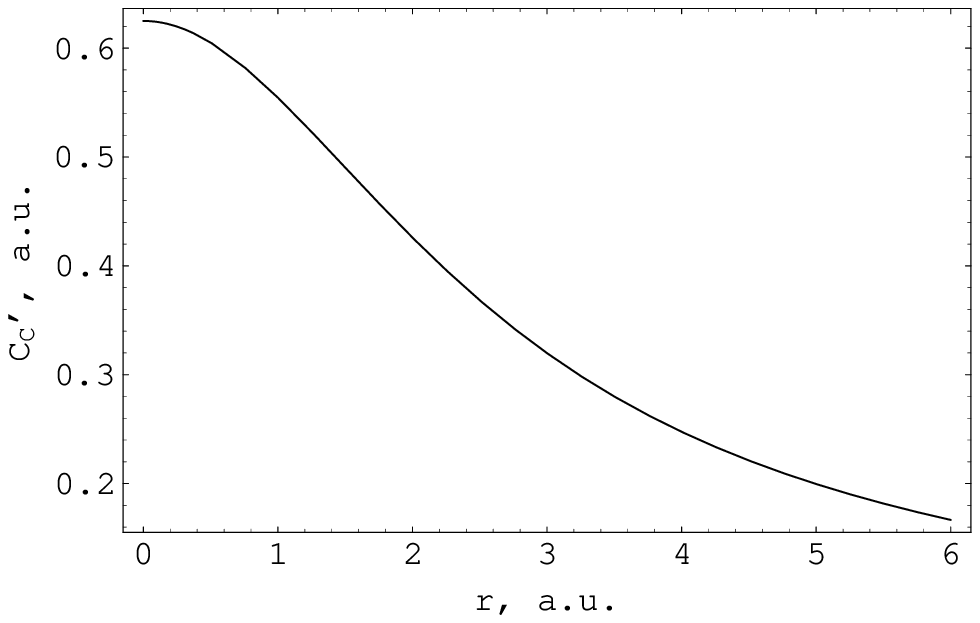}
\caption{The Coloumb integral $\C'_C$ as a function of $\rho$, 
Eq. (\ref{C'0}).}
\label{Fig3}
\end{figure}
\be\l{ExchangeEC}
\E \equiv \E_C= \g(1+\rho)e^{-\rho},
\ee
%
%
\be\l{Sugiura}
\E'_C\equiv\E'_{|V_0=0} = \g
\left(\frac{5}{8}+\frac{23}{20}\rho-\frac{3}{5}\rho^2 -\frac{1}{15}\rho^3
\right)e^{-2\rho} 
+\frac{6\g}{5}\frac{h(\rho)}{\rho},
\ee
\be
h(\rho)= \S^2(\rho)(\ln\rho +C) - \S^2(-\rho)E_1(4\rho)
+2\S(\rho)\S(-\rho)E_1(2\rho),
\ee
%
\begin{figure}
\plot{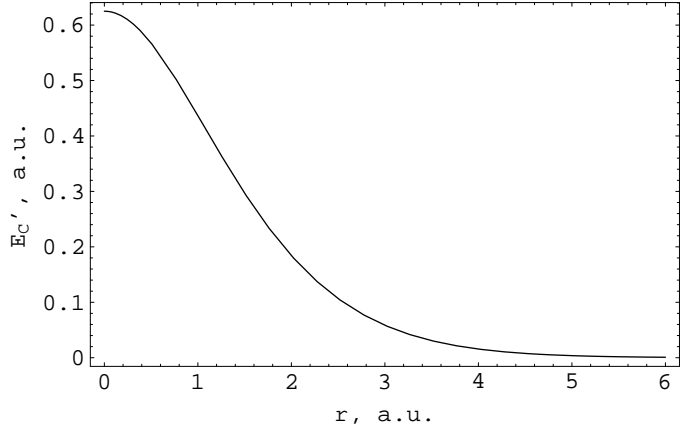}
\caption{The exchange integral $\E'_C$ as a function of $\rho$, 
Eq. (\ref{Sugiura}).}
\label{Fig4}
\end{figure}
\be
E_1(\rho)=\int\limits_{\rho}^{\infty}\frac{e^{-t}}{t}dt,
\ee
\be
\A =  -\frac{1}{2}\g^2 +\g(\g-1), \quad
\A' = -\frac{1}{2}\g^2\S +\g(\g-1)\E,
\ee
where  $C$ is Euler constant, and we have denoted 
\be
\rho = \g R, 
\ee
which can be taken as a second Ritz variational parameter
in addition to $\g$.
The most hard part of work here is the 
exchange integral (\ref{Sugiura}), which was calculated for the first time
by Sugiura (1927), and contains one special function, the exponential 
integral function $E_1(\rho)$.

Our problem is thus to calculate analytically the Hulten potential 
parts of the Coloumb integral (\ref{Coloumb2}) and of the exchange integral 
(\ref{exchange2}), and then vary all the Ritz variational parameters 
in order to minimize the ground state energy (\ref{Emol}),
\be
E_{mol}(\mbox{parameters}) = \mbox{minimum}.
\ee 
In general, we have four parameters in our problem, 
$E_{mol}=E_{mol}(\g, \rho, V_0, r_c)$, with the 
first two parameters characterizing inverse radius of electronic orbit
and the internuclear distance, respectively, and the last two
parameters coming from the Hulten potential. 
However, assuming that the isoelectronium is characterized by 
{\it one} energy level, i.e. $\beta=1$, we have the relation (\ref{r-c})
between $V_0$ and $r_c$ so that we are left with {\it three} 
independent parameters, say, $E_{mol}= E_{mol}(\g, \rho, r_c)$. 
In fact, we have three independent parameters for {\it any} fixed number 
$\beta$ of the levels due to the general relation (\ref{beta}),
\be\l{V0}
V_0 = \frac{\beta^2\hbar^2}{2m r^2_c}, \quad \beta=1,2,\dots.
\ee
Behavior of the energy $E_{mol}$ as a function of $\g$ and $\rho$
is more or less clear owing to known variational analysis of the standard
model of $H_2$ molecule. Namely, $E_{mol}$ reveals a local minimum at some
values of $\g$ and $\rho$. Thus, we should closely analyze 
the $\rc$ dependence of the energy which is specific to 
the Santilli-Shillady model of $H_2$ molecule. 

Below, we turn to the Coloumb integral for the Hulten potential.

\subsubsection{Coloumb integral for Hulten potential}
\l{ColoumbHulten}

To calculate the Hulten part of the Coloumb integral (\ref{Coloumb2})
we use spherical coordinates, $(\rbt,\tt,\pt)$, when integrating over
second electron, and $(\rbo,\tone,\po)$, when integrating over
first electron. 

The integral is
\be
{\cal C}'_h= 4\pi^2\int\limits_{0}^{\pi}
d\tone
\int\limits_{0}^{\infty}\!
d\rbo
\int\limits_{0}^{\pi}\!
d\tt
\int\limits_{0}^{\infty}\!
d\rbt\ 
V_h(r_{12})\Bigl(\frac{\g^3}{\pi}e^{-2\g\rbt}\rbt^2\sin\tt\Bigr)\times
\ee
$$
\times\Bigl(\frac{\g^3}{\pi}e^{-2\g\sqrt{\rbo^2+R^2-\rbo^2R\cos\tone}}
\rbo^2\sin\tone\Bigr),
$$
where Hulten potential is
\be\l{V_h}
V_h(r_{12}) = V_0\frac{e^{-\sqrt{\rbt^2+\rbo^2-2\rbt\rbo\cos\tt}/r_c}}
{1-e^{-\sqrt{\rbt^2+\rbo^2-2\rbt\rbo\cos\tt}/r_c}}.
\ee
Here, we have used 
$$\rao=\sqrt{\rbo^2+R^2-\rbo^2R\cos\tone},$$
$$r_{12}=\sqrt{\rbt^2+\rbo^2-2\rbt\rbo\cos\tt},$$ 
and the fact that integrals over azimuthal angles $\po$ and $\pt$ 
give us $4\pi^2$.

First, we integrate over coordinates of second electron,
\be
I=2\pi\int\limits_{0}^{\pi}\!
d\tt
\int\limits_{0}^{\infty}\!
d\rbt\ 
V_h(r_{12})\Bigl(\frac{\g^3}{\pi}e^{-2\g\rbt}\rbt^2\sin\tt\Bigr).
\ee
Integration over $\tt$ gives us 
\be\l{Ione}
I=\int\limits_{0}^{\infty}\! d\rbt\ (I_1+I_2+I_3+I_4+I_5),
\ee
where
\be
I_1=-4\g^3e^{-2\g\rbt}\rbt^2,
\ee
\be
I_2=-2\g^3\rc\frac{\rbt}{\rbo}\sqrt{(\rbo-\rbt)^2} 
    e^{-2\g\rbt}\ln(1-e^{\sqrt{(\rbo-\rbt)^2}/\rc}),
\ee
\be
I_3=-2\g^3\rc\frac{\rbt}{\rbo}\sqrt{(\rbo+\rbt)^2} 
    e^{-2\g\rbt}\ln(1-e^{\sqrt{(\rbo+\rbt)^2}/\rc}),
\ee
\be
I_4=2\g^3\rc^2\frac{\rbt}{\rbo}
    e^{-2\g\rbt}Li_2(e^{\sqrt{(\rbo-\rbt)^2}/\rc}),
\ee
\be
I_5=2\g^3\rc^2\frac{\rbt}{\rbo}
    e^{-2\g\rbt}Li_2(e^{\sqrt{(\rbo+\rbt)^2}/\rc}),
\ee
and
\be
Li_2(z) = \sum\limits_{k=1}^{\infty}\frac{z^k}{k^2}
        = \int\limits_{z}^{0}\frac{\ln(1-t)}{t}dt 
\ee
is dilogarithm function.

Now, we turn to integrating over $\rbt$. For $I_1$ we have
\be
\int\limits_{0}^{\infty}\! d\rbt\  I_1 = -1.
\ee
In $I_2$, we should keep $(\rbo-\rbt)$ to be positive
so we write down two separate terms,
\be
\int\limits_{0}^{\infty}\! d\rbt  I_2 =
I_{21}+I_{22}\equiv  
\int\limits_{0}^{\rbo}\! d\rbt\   I_2(\rbt<\rbo) +
\int\limits_{\rbo}^{\infty}\! d\rbt\   I_2(\rbt>\rbo).
\ee
In these two integrals, $I_{21}$ and $I_{22}$, we change variable
$\rbt$ to $x$ and $y$, respectively,
\be
x=(\rbo-\rbt)/\rc, \quad \rbo/\rc<x<0, \quad
y=(\rbt-\rbo)/\rc, \quad 0<y<\infty,
\ee
in order to simplify integrating.
In terms of these variables, we have
\be
I_{21} = \int\limits_{\rbo/\rc}^{0}\! dx\ 
2\g^3\rc^3(x-\frac{\rc}{\rbo}x^2)e^{-2\g(\rbo-\rc x)}\ln(1-e^x),
\ee
\be
I_{22} = -\int\limits_{0}^{\infty}\! dy\ 
2\g^3\rc^3(y+\frac{\rc}{\rbo}y^2)e^{-2\g(\rbo+\rc y)}\ln(1-e^y).
\ee
We are unable to perform these integrals directly.
To calculate these integrals we use method of differentiating in 
parameter. Namely, we use simpler integrals,
\be
L_1=\int dx\ e^{2\g\rc x}\ln(1-e^x)
\ee 
and
\be
L_2=\int dy\ e^{-2\g\rc y}\ln(1-e^y),
\ee 
and differentiate them in parameter $\rc$
to reproduce $I_{21}$ and $I_{22}$. (One can use parameter $\g$
for this purpose, or introduce an independent parameter
putting it to one after making calculations, with the same result.)
Namely, by using definitions of 
$L_1$ and $L_2$ we have
\be\l{I21}
I_{21}= 2\g^3\rc^3(\frac{1}{2}\frac{d}{d\rc}L_1
-\frac{\rc}{4\rbo}\frac{d^2}{d\rc^2}L_1)\ |_{x=\rbo/\rc}^{x=0},
\ee
\be\l{I22}
I_{22} = -
2\g^3\rc^3(-\frac{1}{2}\frac{d}{d\rc}L_2
-\frac{\rc}{4\rbo}\frac{d^2}{d\rc^2}L_2)\ |_{y=0}^{y=\infty}.
\ee
Now, the problem is to calculate indefinite integrals,  
$L_1$ and $L_2$, which make basis for further algebraic calculations. 
After making the calculations, we have 
\be\l{L1}
L_1 = \frac{1}{4\g^2\rc^2}e^{2\g\rc}
\Bigl(2\g\rc(\Phi(e^x,1,2\g\rc)+\ln(1-e^x))-1\Bigr)
\ee
and 
\be\l{L2}
L_2 = -\frac{1}{4\g^2\rc^2}e^{-2\g\rc}
\Bigl(2\g\rc(\Phi(e^y,1,-2\g\rc)+\ln(1-e^y))+1\Bigr),
\ee
where
\be\l{Lerch}
\Phi(z,s,a) = \sum\limits_{k=0}^{\infty}\frac{z^k}{(a+k)^s}, \quad
a+k\not=0,               
\ee
is {\it Lerch function}, which is a generalization of polylogarithm function
$Li_n(z)$ and Riemann $\zeta$-function. Particularly, 
$Li_2(z)=\Phi(z,2,0)$.
Also, we note that the Lerch function arises when dealing with Fermi-Dirac
distribution, e.g.,
\be
\int\limits_{0}^{\infty}\! dk\ \frac{k^s}{e^{k-\mu}+1}
=e^\mu\Gamma(s+1)\Phi(-e^\mu,s+1,1).
\ee
Below, we will need in derivatives of Lerch function $\Phi(z,s,a)$ 
in third argument. By using the definition (\ref{Lerch}) we obtain 
directly
\be\l{der1}
\frac{d}{da}\Phi(z,s,a) \equiv \Phi'(z,s,a) = -s\Phi(z,s+1,a),
\ee
\be\l{der2}
\frac{d^2}{da^2}\Phi(z,s,a) \equiv \Phi''(z,s,a) = s(s+1)\Phi(z,s+2,a).
\ee
Inserting (\ref{L1}) and (\ref{L2}) into (\ref{I21}) and (\ref{I22})
we get
\be\l{I21'}
I_{21} = \frac{1}{4\g\rbo}\Bigl(e^{-2\g(\rbo-\rc x)}
(3+2\g(\rbo-(2+\g\rbo)\rc x+\g\rc^2x^2)-
\ee
$$
-2\g\rc((1+\g(\rbo-2(1+\g\rbo)\rc x+2\g\rc^2x^2))
[\Phi(e^x,1,2\g\rc)+ \ln(1-e^x)]+
$$
$$
+2\g\rc(-(1+\g(\rbo-2\rc x))\Phi'(e^x,1,2\g\rc)
+\g\rc\Phi''(e^x,1,2\g\rc)))\Bigr)\ |_{x=\rbo/\rc}^{x=0},
$$
\be\l{I22'}
I_{22} = \frac{1}{4\g\rbo}\Bigl(e^{-2\g(\rbo+\rc x)}
(3+2\g(\rbo+(2+\g\rbo)\rc x+\g\rc^2x^2)+
\ee
$$
+2\g\rc((1+\g(\rbo+2(1+\g\rbo)\rc x+2\g\rc^2x^2))
[\Phi(e^x,1,-2\g\rc)+\ln(1-e^x)] +
$$
$$
+2\g\rc((1+\g(\rbo+2\rc x))\Phi'(e^x,1,-2\g\rc)
+\g\rc\Phi''(e^x,1,-2\g\rc)))\Bigr)\ |_{y=0}^{y=\infty}.
$$
Now, we have to use the above derivatives (\ref{der1}) and (\ref{der2})
of Lerch function to obtain final expressions for $I_{21}$ and $I_{22}$. 
Then, we should take the limits $x\to \rbo/\rc$, $x\to 0$, 
and $y\to 0$, $y\to\infty$, respectively.
The endpoints $x=\rbo/\rc$ and $y=\infty$ can be inserted 
easily, with the endpoint $y=\infty$ yielding zero, while the limits 
$x \to 0$ and $y\to 0$ require some care because of the presence of 
some divergent terms.

To collect all the terms, we sum up $I_{21}$ and $(-1)I_{22}$ given 
by (\ref{I21'}) and (\ref{I22'}), put $x=y$, and take common limit
$x\to 0$, inserting $x=0$ for polynomial and exponential (welldefined) 
terms. We get
$$
I_{21}-I_{22}|_{x\to 0}=
$$
\be
=-\frac{1}{2\rbo}\Bigl(\rc e^{-2\g\rbo}
(2\g\rc(1+\g\rbo)[\Phi(e^x,2,2\g\rc)-\Phi(e^x,2,-2\g\rc)]+
\ee
$$
+4\g^2\rc^2[\Phi(e^x,3,2\g\rc)+\Phi(e^x,3,-2\g\rc)]+
$$
$$
+(1+\g\rbo)[\Phi(e^x,1,2\g\rc)+\Phi(e^x,1,-2\g\rc)-2\ln(1-e^x)]
\Bigr)|_{x\to 0}
$$
The limits of Lerch functions of second, $\Phi(e^x,2,\pm 2\g\rc)$, and 
third,\\
$\Phi(e^x,3,\pm 2\g\rc)$, order, at $x\to 0$, are welldefined while 
each of the terms in 
\be\l{B}
B(2\g\rc)\equiv
\lim_{x\to 0}[\Phi(e^x,1,2\g\rc)+\Phi(e^x,1,-2\g\rc)-2\ln(1-e^x)]
\ee
is {\it divergent} since Lerch function of first order, 
$\Phi(e^x,1,\pm 2\g\rc)$, increases unboundedly at $x\to 0$. 
We will analyze this limit below, to identify the condition
at which the divergencies cancel each other. Now, we collect
all the terms obtaining final result for the integral 
in the form
\be\l{I2}
\int\limits_{0}^{\infty}\! d\rbt\ I_2 = \frac{1}{4\rbo}\Bigl(
\frac{1}{\g}(-3+2\g\rbo+8\g^3\rc^3\Phi(e^{\rbo/\rc},3,2\g\rc)+
\ee
$$
+2\g\rc(1-\g\rc)
[\Phi(e^{\rbo/\rc},1,2\g\rc)+ 2\g\rc\Phi(e^{\rbo/\rc},2,2\g\rc)
+\ln(1-e^{\rbo/\rc})])-
$$
$$
-2\rc e^{-2\g\rbo}(1+\g\rbo)\bigl\{
B(2\g\rc)+2\g\rc[\zeta(2,2\g\rc)-\zeta(2,-2\g\rc)]+
$$
$$
+4\g^2\rc^2[\zeta(3,2\g\rc)+\zeta(3,-2\g\rc)]\bigr\}
\Bigr),
$$
where
\be
\zeta(s,a)=\sum\limits_{k=1}^{\infty}\frac{1}{(a+k)^s}, \quad a+k\not=0,
\ee
is generalized Riemann $\zeta$-function. 
The values of $\zeta(2,\pm 2\g\rc)$ and $\zeta(3,\pm 2\g\rc)$
entering (\ref{I2}) are welldefined. 
For example, at $\g=1.4$  and $\rc=0.0048$, we have
\be
\zeta(2,\pm 2\g\rc) \simeq 5537, \quad \zeta(3,\pm 2\g\rc) \simeq 2462.
\ee

Now, we turn to close consideration of the limit (\ref{B})
entering (\ref{I2}).
Let us calculate it for the particular value $2\g\rc=1/100$.
Using expansion of each term of $B$ around $x=0$, we obtain
\be\l{B100}
B(\frac{1}{100}) = \lim_{s\to 1}\Bigl[100 
-\frac{1}{\Gamma(\frac{1}{100})}\bigl\{
100\Gamma(\frac{101}{100})(C+\ln(1-s)+\psi(\frac{1}{100}))
\bigr\}-
\ee
$$
-\frac{1}{99\Gamma(\frac{99}{100})}\bigl\{
100\Gamma(\frac{199}{100})(C+\ln(1-s)+\psi(\frac{99}{100}))
\bigr\}
+2\ln(1-s)+ O(1-s)\Bigr],
$$
where we have denoted, for brevity, $s=e^x$,
\be
\psi(z) =\sum\limits_{n=0}^{\infty}\frac{1}{z+n}
=\frac{\Gamma'(z)}{\Gamma(z)} 
\ee
is digamma function, $\Gamma(z)$ is Euler gamma function, 
and $C$ is Euler constant.
Using elementary properties of gamma function we obtain from Eq.(\ref{B100})
\be\l{B100OK}
B(\frac{1}{100}) = 100-2C -\psi(\frac{1}{100})-\psi(\frac{99}{100}),
\ee
so one can see that the logarithmic divergent terms 
cancel each other, and the limit is welldefined for 
$2\g\rc=1/100$.
The same is true for any {\it integer} value of
\be\l{k}
k=\frac{1}{2\g\rc}
\ee	
while at noninteger $k$ the limit $B(\frac{1}{k})$ blows up. 
Generalizing the above particular result (\ref{B100OK}), 
we can write down
\be\l{Bk}
B(\frac{1}{k}) = k-2C -\psi(\frac{1}{k})-\psi(1-\frac{1}{k}),
\ee
for any integer $k>2$.

This highly remarkable result means that to have finite 
value of the Coloumb integral we should use the condition that 
$\lambda^{-1}\equiv (2\g\rc)^{-1}=k$ is an integer number. 
Recalling that typically $\g \simeq 1.5$ and $\rc \simeq 0.01$ 
we have the integer number $k\simeq 30$.

Now, we turn to the next integral, $I_3$. It is similar to $I_2$ so that 
we present the final expression,
\be\l{I3}
\int\limits_{0}^{\infty}\! d\rbt\ I_3 =
\frac{1}{4\g\rbt}
\Bigl(
3
+2\g\rbo 
+2\g\rc(1+\g\rbo)[\Phi(e^{\rbo/\rc},1,-2\g\rc)+\ln(1-e^{\rbo/\rc})]
\ee
$$
-4\g^2\rc^2(1+\g\rbo)\Phi(e^{\rbo/\rc},2,-2\g\rc)
+8\g^3\rc^3\Phi(e^{\rbo/\rc},3,-2\g\rc)
\Bigr).
$$

The integral $I_4$ is more complicated, 
\be
\int\limits_{0}^{\infty}\! d\rbt\ I_4=I_{41}+I_{42},
\ee
where
\be
I_{41}=\int\limits_{0}^{\rbo}d\rbt 
2\g^3\rc^2\frac{\rbt}{\rbo}
e^{-2\g\rbt}Li_2(e^{(\rbo-\rbt)/\rc}),
\ee
\be
I_{42}=\int\limits_{\rbo}^{\infty}d\rbt 
2\g^3\rc^2\frac{\rbt}{\rbo}
e^{-2\g\rbt}Li_2(e^{(\rbt-\rbo)/\rc}).
\ee
Introducing variables
\be
x=(\rbo-\rbt)/\rc, \quad
y=(\rbt-\rbo)/\rc, 
\ee
we rewrite the integrals in the form
\be
I_{41}=\int\limits_{\rbo/\rc}^{0}dx\ 
2\g^3\rc^3e^{2\g(\rbo-\rc x)}
[Li_2(e^x)-\frac{\rc}{\rbo}x Li_2(e^x)],
\ee
\be
I_{42}=-\int\limits_{0}^{\infty}dy\ 
2\g^3\rc^3e^{2\g(\rbo+\rc y)}
[Li_2(e^y)+\frac{\rc}{\rbo}y Li_2(e^y)].
\ee
In the r.h.s. of $I_{41}$, the first term can be calculated 
directly in terms of Lerch function while the second term can be 
obtained from the first term by differentiating it in 
the parameter, for which we choose again $\rc$.
Namely, the basic integral, which we will use to calculate $I_{41}$, is 
\be\l{M1}
M_1=\int\limits_{x_0}^{0}dx\ 
e^{2\g\rc x}Li_2(e^x),
\ee
for which we have
\be\l{M1calc}
M_1=\frac{1}{24\g^3\rc^3\Gamma(2\g\rc)}
\Bigl(
3(e^{2\g\rc x_0}-1)\Gamma(2\g\rc)+
\ee
$$
+\Gamma(1+2\g\rc)(\g\rc\pi^2-3C-3\psi(2\g\rc))-
$$
$$
-3e^{2\g\rc x_0}\Gamma(1+2\g\rc)(\Phi(e^{x_0},1,2\g\rc)+\ln(1-e^{x_0})
+2\g\rc Li_2(e^{x_0}))
\Bigr).
$$
We use this result in the first term of $I_{41}$. Differentiating
$M_1$ given by Eqs. (\ref{M1}) and (\ref{M1calc}) in $\rc$, we 
reproduce the second term of $I_{41}$, up to a factor.
So, collecting these results and inserting $x_0=\rbo/\rc$ we obtain
after some algebra
\be\l{I41calc}
I_{41}=
-\frac{1}{12\rbo\Gamma(1+2\g\rc)}
\Bigl(
e^{-2\g\rc}[\rc(9+4\pi^2\g^3\rc^2\rbo)\Gamma(2\g\rc)+
\ee
$$
\Gamma(1+2\g\rc)(6C\rc-3\rbo-6C\g\rc\rbo- \pi^2\g\rc^2
-6(\g\rbo-1)\rc\psi(2\g\rc)-
$$
$$
-6\g\rc^2\psi'(2\g\rc))]+3(2\rbo\Gamma(1+2\g\rc)-3\rc\Gamma(2\g\rc)+
$$
$$
+2\rc\Gamma(1+2\g\rc)
[(1-2\g\rbo)\Phi(e^{\rbo/\rc},1,2\g\rc)+\g\rc\Phi(e^{\rbo/\rc},2,2\g\rc)+
$$
$$
+(1-2\g\rbo)\ln(1-e^{\rbo/\rc})+\g(1-4\g\rbo)\rc\psi'(e^{\rbo/\rc})]
\Bigr),
$$
where $\psi'(z)=d\psi(z)/dz$ is derivative of digamma function.

To calculate $I_{42}$ we use a similar method. However, care should be 
exerted when taking limit $y \to 0$. The basic integral, which we will use
to calculate $I_{42}$, is 
\be
M_2=-\int dy\ e^{2\g\rc y}Li_2(e^{-y}),
\ee
where we have replaced $y\to -y$ so that the endpoints will be
due to $0<y<-\infty$. The result for $M_2$ is 
\be
M_2 = \frac{1}{8\g^3\rc^3}
e^{2\g\rc y}(1+2\g\rc e^y\Phi(e^y,1,1+2\g\rc)
+2\g\rc\ln(1-e^{-y}) - 
\ee
$$
-4\g^2\rc^2Li_2(e^{-y})).
$$
We should insert here the endpoints $y=0$ and $y=-\infty$.
In the limit $y\to -\infty$, $M_2$ is zero. 
In the limit $y\to 0$, we have
\be
Li_2(e^{-y})|_{y\to0} = \frac{\pi^2}{6}
\ee
and, assuming that $k=1/(2\g\rc)$ is an integer number,
\be
\Phi(e^y,1,1+2\g\rc)+\ln(1-e^{-y})|_{y\to0} 
= -(\frac{1}{2\g\rc}+C+\psi(2\g\rc)).
\ee
Thus, 
\be
M_2|_{y=0}^{y=-\infty} 
= \frac{1}{12\g^2\rc^2}(3C+\pi^2\g\rc+3\psi(2\g\rc)),
\ee
for integer $k$. We should point out that, 
in the case of noninteger $k$, $M_2$ increases unboundedly at $y\to 0$. 
Using this result in $I_{42}$, we obtain
\be\l{I42calc}
I_{42}= 
-\frac{\rc}{12\rbo}e^{-2\g\rbo}\Bigl( 6C(1+\g\rbo)+\pi^2\g\rc
+2\pi^2\g^2\rc\rbo+
\ee
$$
+6(1+\g\rbo)\psi(2\g\rc)-6\g\rc\psi'(2\g\rc)\Bigr).
$$
Summing up $I_{41}$ given by (\ref{I41calc}) and $I_{42}$
given by (\ref{I42calc}), we get
\be
\int\limits_{0}^{\infty}\! d\rbt\ I_4= 
=\frac{1}{4\rbo\Gamma(1+2\g\rc)}
\Bigl(
3\rc\Gamma(2\g\rc)+e^{-2\g\rc}[-\rbo\Gamma(1+2\g\rc)+
\ee
$$
+\rc(-3\Gamma(2\g\rc)+4\Gamma(1+2\g\rc)(-(1+\g\rbo)(C+\psi(2\g\rc)
+\g\rc\psi'(2\g\rc)))]-
$$
$$
-2\rc\Gamma(1+2\g\rc)[\Phi(e^{\rbo/\rc},1,2\g\rc)+\ln(1-e^{\rbo/\rc})
+\g\rc(\Phi(e^{\rbo/\rc},2,2\g\rc)+
$$
$$
+Li_2(e^{\rbo/\rc}))]
\Bigr).
$$

The integral $I_5$ is similar to $I_4$ so that we present the final
expression,
\be
\int\limits_{0}^{\infty}\! d\rbt\ I_5 
= -\frac{1}{8\g\rbo}
\Bigl(
3+4\g\rc[
e^{-\rbo/\rc}(\Phi(e^{-\rbo/\rc},1,1+2\g\rc)+
\ee
$$
+\Phi(e^{-\rbo/\rc},2,1+2\g\rc))
+\ln(1-e^{\rbo/\rc})-\g\rc\psi'(e^{\rbo/\rc})]
\Bigr).
$$

Now, we are in a position to 
sum up all the calculated integrals $I_1,\dots,I_5$, and obtain, due to 
(\ref{Ione}), the following final expression
for the Coloumb integral over coordinates of {\it second} electron,
\be\l{Icalc1}
I(\rbo)=
-(\frac{1}{2}+\frac{5}{8\g\rbo})e^{-2\g\rbo}+
\ee
$$
+\frac{1}{2}\g\rc
\Bigl[
\pi(1+\frac{1}{\g\rbo})\mbox{ctg}(2\g\rc\pi)e^{-2\g\rbo}
-\frac{1}{\g\rbo}e^{-\rbo/\rc}\Phi(e^{-\rbo/\rc},1,1+2\g\rc)+
$$
$$
+\Phi(e^{\rbo/\rc},1,-2\g\rc)-\Phi(e^{\rbo/\rc},1,2\g\rc)
+\frac{1}{\g\rbo}\Phi(e^{\rbo/\rc},1,-2\g\rc)
\Bigr]+
$$
$$
+\g^2\rc^2
\Bigl[
-\frac{1}{2\g\rbo}e^{-\rbo/\rc}\Phi(e^{-\rbo/\rc},2,-2\g\rc)
-\Phi(e^{\rbo/\rc},2,-2\g\rc)-
$$
$$
-\Phi(e^{\rbo/\rc},2,2\g\rc)
+\frac{1}{\g\rbo}
(\frac{1}{2}\Phi(e^{\rbo/\rc},2,2\g\rc)-\Phi(e^{\rbo/\rc},2,-2\g\rc))+
$$
$$
+\frac{1}{\g\rbo}e^{-2\g\rbo}\psi'(2\g\rc)
+e^{-2\g\rbo}
(1+\frac{1}{\g\rbo})(\zeta(2,-2\g\rc)-\zeta(2,2\g\rc))
\Bigr]+
$$
$$
+\frac{2}{\g\rbo}\g^3\rc^3
\Bigl[
\Phi(e^{\rbo/\rc},3,2\g\rc)+\Phi(e^{\rbo/\rc},3,-2\g\rc)-
$$
$$
-e^{-2\g\rbo}(\zeta(3,2\g\rc)+\zeta(3,-2\g\rc))
\Bigr],
$$
where we have collected the terms due to power degrees of $\rc$.
It should be stressed that here $(2\g\rc)^{-1}$ is assumed to be 
an integer number. The above expression represents the Hulten part of 
the electrostatic potential caused by charge distribution of 
the second electron.

Next step is to integrate (\ref{Icalc1}) over the coordinates
of {\it first} electron, 
\be\l{CH1}
{\cal C}'_h=2\pi\int\limits_{0}^{\pi}d\tone\int\limits_{0}^{\infty}d\rbo\ 
I(\rbo) \frac{\g^3}{\pi}e^{-2\g\sqrt{\rbo^2+R^2-\rbo^2R\cos\tone}}
\rbo^2\sin\tone.
\ee
Prior to that, we denote 
\be
\lambda=2\g\rc=\frac{1}{k}, \quad
r=\g\rbo, 
\ee
and rewrite Eq. (\ref{Icalc1}) in a more compact form,
\be\l{Icalc}
I(r)=
-(\frac{1}{2}+\frac{5}{8r})e^{-2r}
+\frac{1}{4}\lambda
\Bigl[
\pi(1+\frac{1}{r})\mbox{ctg}(\pi\lambda)e^{-2r}
-\frac{1}{r}e^{-2r/\lambda}\Phi(e^{-2r/\lambda},1,1+\lambda)+
\ee
$$
+\Phi(e^{2r/\lambda},1,-\lambda)-\Phi(e^{2r/\lambda},1,\lambda)
+\frac{1}{r}\Phi(e^{2r/\lambda},1,-\lambda)
\Bigr]+
$$
$$
+\frac{\lambda^2}{4}
\Bigl[
-\frac{1}{2r}e^{-2r/\lambda}\Phi(e^{-2r/\lambda},2,-\lambda)
-\Phi(e^{2r/\lambda},2,-\lambda)-\Phi(e^{2r/\lambda},2,\lambda)+
$$
$$
\frac{1}{r}
(\frac{1}{2}\Phi(e^{2r/\lambda},2,\lambda)-\Phi(e^{2r/\lambda},2,-\lambda)
+e^{-2r}\psi'(\lambda))
+e^{-2r}
(1+\frac{1}{r})(\zeta(2,-\lambda)-\zeta(2,\lambda))
\Bigr]
$$
$$
+\frac{\lambda^3}{4r}
\Bigl[
\Phi(e^{2r/\lambda},3,\lambda)+\Phi(e^{2r/\lambda},3,-\lambda)
-e^{-2r}(\zeta(3,\lambda)+\zeta(3,-\lambda))
\Bigr].
$$
Since $I(r)$ does not depend on $\tone$ one can easily integrate 
over $\tone$ in Eq.(\ref{CH1}), and then change variable
$\rbo$ to $r=\g\rbo$, obtaining
\be\l{CH2}
{\cal C}'_h=
\frac{1}{2\rho}\int\limits_{0}^{\infty}dr\  
I(r) \Bigl[(1+2\sqrt{(\rho-r)^2})re^{-2\sqrt{(\rho-r)^2}}
-(1+2(\rho+r))re^{-2(\rho+r)}\Bigr],
\ee
where $\rho=\g R$. Again, we should use separate intervals 
to keep $(\rho-r)$ to be positive, namely, 
we rewrite ${\cal C}'_h$ as
\be\l{CH3}
{\cal C}'_h= J_1+J_2+J_3,
\ee
where
\be
J_1=\frac{1}{2\rho}\int\limits_{0}^{\rho}dr\  I(r) 
(1+2\rho-2r)re^{-2(\rho-r)},
\ee
\be
J_2=\frac{1}{2\rho}\int\limits_{\rho}^{\infty}dr\  I(r) 
(1+2r-2\rho)re^{-2(r-\rho)},
\ee
\be
J_3=-\frac{1}{2\rho}\int\limits_{0}^{\infty}dr\  I(r) 
(1+2\rho+2r)re^{-2(\rho+r)}.
\ee
Now, we are ready to make integration over the last remaining variable, $r$,
to obtain complete analytical expression of the Coloumb integral for the 
Hulten potential. 

However, Lerch functions entering Eq.(\ref{Icalc})
make obstacle to do integral (\ref{CH3}) for a general case because 
they have different functional form for different values of the parameter 
$\lambda$. So, each of the above integrals $J_{1,2,3}$ 
should be calculated independently for 
every numerical value of $\lambda$. 
Moreover, for the values of interest, e.g., 
$\lambda= 1/30$, {\it each} Lerch function is expressed in the form 
of sum of elementary functions with too big number of nontrivial terms 
to handle them (incomplete Euler beta function arises here). 
So, the integral cannot be reliably 
calculated even for a single value of $\lambda$, within the interval 
of interest, $\lambda= 1/30, 1/31,\dots, 1/100$.
Also, elementary analysis shows that we can not implement the assumption
of small $\rc$ into Eq.(\ref{Icalc}), to use first order approximation
in $\rc$. Indeed, Lerch functions in (\ref{Icalc}) contain $\rc$ 
both in first and third argument so that their asymptotics at $\rc\to0$
make no sense.

Thus, we stop here further calculation of the Coloumb integral 
${\cal C}'_h$ getting, however, as our main result the fact that 
$(2\g\rc)^{-1}$ should be integer number, in the variational approach 
to the model, to have finite energy of the ground state. 
We consider this as very interesting result deserving rather involved 
calculations made above. 

Also, we have a detailed technical view on the problems which arise when 
dealing with molecular integrals with the Hulten potential. 
Practically, this means that there is
a very little hope that the {\it exchange} integral (\ref{exchange2}), 
which is structurally much more complicated than the above considered 
Coloumb one, can be calculated exactly for the case of Hulten potential.

Because of these difficulties, below we use appropriate {\it simplified} 
potentials, instead of Hulten potential,
to have some analytical set up for the variational analysis of the 
Santilli-Shillady model. Clearly, by this we go to some approximation 
to the original Santilli-Shillady model.

\subsubsection{Coloumb integral for exponential screened Coloumb potential}
\l{ColoumbExp}

We use simple function to mimic Hulten potential. 
Namely, we approximate the general potential (\ref{V12}) by
\be\l{VExp}
V(r_{12}) = V_C+V_e=\frac{e^2}{r_{12}}-\frac{Ae^{-r_{12}/r_c}}{r_{12}},
\ee
where $A$ and $r_c$ are positive parameters.
It has similar behavior both at short and long distances. 
Indeed, at long distances, $r_{12}\to\infty$, we can ignore $V_e$ 
and the behavior is solely due to the Coloumb potential while 
its series expansion about the point $r_{12}=0$ (short distances) is 
\be
V(r_{12})_{|r_{12}\to 0} = 
\frac{e^2-A}{r_{12}} + \frac{A}{r_c}-\frac{A}{2r_c}r_{12} + O(r^2_{12}).
\ee
Here, we should put 
$A=V_0r_c$
to have the same coefficient at $r^{-1}_{12}$ in the $r_{12}\to 0$ 
asymptotics as it is in the case of Hulten potential; see Eq.(\ref{series}).
Using Eq.(\ref{V0}) we have
\be\l{V0rc}
A=V_0r_c = \frac{\beta^2\hbar^2}{2m r_c}, \quad \beta=1,2,\dots,
\ee
where $\beta$ is a number of energy levels of isoelectronium. 
Taking $\beta=1$ we have, in atomic units ($\hbar=1$, $m=m_e/2=1/2$), 
\be\l{Arc}
A= \frac{1}{r_c}.
\ee

Below, we calculate the Coloumb integral
(\ref{Coloumb2}), with the exponential screened
Coloumb potential $V_e$ defined by Eq.(\ref{VExp}),
\be\l{Coloumb3}
\C'_{E} = \int dv_1dv_2\ 
\left(\frac{e^2}{r_{12}}-\frac{Ae^{-r_{12}/r_c}}{r_{12}}\right)
|f(r_{a1})|^2|f(r_{b2})|^2,
\ee
Below, we present some details of calculation of 
the Coloumb integral (\ref{Coloumb3}). Apart from the case 
of Hulten potential considered in Sec.~\ref{ColoumbHulten}, 
it appears that this integral can be 
calculated in terms of elementary functions.

The integral we are calculating is
\be
\C'_{e}=\int dv_1dv_2\ \frac{Ae^{-\r/\rc}}{\r} |f(\rao)|^2|f(\rbt)|^2,
\ee
where
\be
f(r)=\sqrt{\frac{\g^3}{\pi}}e^{-\g r},
\ee
and $dv_1$ and $dv_2$ are volume elements for the first and second electron,
respectively. We use spherical coordinates.
In spherical coordinates $(\rbt, \tt, \pt)$, with polar axis directed
along the vector $\vec r_{b1}$, we have
\be\l{r12app}
\r = \sqrt{\rbo^2+\rbt^2-2\rbo\rbt\cos\tt}.
\ee
We use these coordinates when integrating over second electron.
In spherical coordinates $(\rbo, \tone, \po)$, with polar axis 
directed along the vector $\vec R$, we have
\be\l{ra1app}
\rao = \sqrt{\rbo^2+R^2-2\rbo R\cos\tone}.
\ee
We use these coordinates when integrating over first electron.

First, we integrate over angular coordinates of second electron,
\be
I_1 = \int\limits_{0}^{2\pi}d\pt\int\limits_{0}^{\pi}d\tt \ 
\frac{Ae^{-\r/\rc}}{\r}\frac{\g^3}{\pi}e^{-2\g\rbt}\rbt^2\sin\tt,
\ee
where $\r$ is defined by (\ref{r12app}). It is relatively easy
to calculate this integral,
\be
I_1 = \frac{2A\g^3\rc}{\rbo}e^{-2\g\rbt}
\left(
e^{-\sqrt{(\rbt-\rbo)^2}/\rc}-e^{-\sqrt{(\rbt+\rbo)^2}/\rc}
\right).
\ee
Further, integrating on radial coordinate $\rbt$ must be 
performed in separate intervals,
\be
I_2 =\int\limits_{0}^{\rbo}d\rbt\ I_1(\rbt<\rbo)+
\int\limits_{\rbo}^{\infty}d\rbt\ I_1(\rbt>\rbo),
\ee
where 
\be
\sqrt{(\rbo-\rbt)^2} = 
\left\{ 
\begin{array}{cc}
r_{b1}-r_{b2}, & \quad r_{b2}<r_{b1},\\
r_{b2}-r_{b1}, & \quad r_{b2}>r_{b1},
\end{array}
\right.
\ee
with the result
\be
I_2= \frac{4A\g^3\rc^2(4\g\rc^2(e^{-\rbo/\rc}-e^{-2\g\rbo})
+\rbo e^{-2\g\rbo}(1-4\g^2\rc^2))}
{\rbo(1-4\g^2\rc^2)^2}.
\ee
Now, we turn to integrating over coordinates of first electron,
$(\rbo, \tone, \po)$,
\be
I_3 =  \int\limits_{0}^{\pi}d\tone\int\limits_{0}^{2\pi} d\po\ 
I_2 \frac{\g^3}{\pi}e^{-2\g\rao}\rbo^2\sin\tone,
\ee
where $\rao$ is defined by (\ref{ra1app}). We obtain
after tedious calculations
\be
I_3=\frac{2A\g^4\rc^2}{R(1-4\g^2\rc^2)^2}
e^{-2\g(\sqrt{(R-\rbo)^2}+\rbo+\sqrt{(R+\rbo)^2})-\rbo/\rc}
\ee
$$
\times\left[
e^{2\g\sqrt{(R+\rbo)^2}}
-2\g e^{2\g\sqrt{(R+\rbo)^2}}\sqrt{(R-\rbo)^2}
+2\g e^{2\g\sqrt{(R-\rbo)^2}}\sqrt{(R+\rbo)^2}
\right]
$$
$$
\times\left[
e^{\rbo/\rc}(4\g(1+\g\rbo)\rc^2 - \rbo) - 4\g\rc^2e^{2\g\rbo}
\right].
$$
Again, we must further integrate in $\rbo$ by separate intervals,
\be\l{I4}
I_4 =\int\limits_{0}^{R}d\rbo\ I_3(\rbo<R)+
\int\limits_{R}^{\infty}d\rbo\ I_3(\rbo>R)
\equiv I_{41}+I_{42},
\ee
obtaining after rather tedious calculations
\be
I_{41}= \frac{Ae^{-6\g R}\g^2\rc^2}{12R(1-4\g^2\rc^2)^4}\times
\ee
$$
\Bigl[
96e^{2\g R -R/\rc}\g^3
\Bigl(    (4\g(2\g R\rc+R+\rc)+1)(1-2\g\rc)^2 
           +e^{4\g R}(2\g\rc+1)^2(4\g\rc-1) 
\Bigr)\rc^3
$$
$$
-3e^{4\g R}
+3\g
\Bigl(  -64\g^5(\g R( 8\g R+13)+4)\rc^6
        +16\g^3(\g R(24\g R+31)+9)\rc^4
$$
$$
        -4 \g  (\g R(24\g R+23)+6)\rc^2  
        +R(8\g R+5)
\Bigr) -e^{4\g R}\g\times
$$
$$
\Bigl(  64\g^5(\g R(4\g R(2\g R+9)+57)+36)\rc^6
       -48\g^3(\g R(4\g R(2\g R+7)+21)- 9)\rc^4
$$
$$
       +12\g  (\g R(4\g R(2\g R+5)+ 1)- 6)\rc^2
       -R    (4\g R(2\g R+3)-3)+3) 
\Bigr)
+3
\Bigr],
$$
\be
I_{42} = -\frac{Ae^{-6\g R}\g^2\rc^2}{4R(1-2\g\rc)^2(1+2\g\rc)^4}\times
\ee
$$
\times\Bigl[
1-e^{4\g R}-128\g^6 R^2\rc^5 +\g\Bigl((5-3e^{4\g R})R-4(e^{4\g R}-1)\rc\Bigr)+
$$
$$
+16\g^5 R \rc^3
\Bigl(-8R+ (16e^{2\g R-R/\rc}+3e^{4\g R}-13)\rc\Bigr)+
$$
$$
+16e^{-R/\rc}\g^4\rc^3
\Bigl((8e^{2\g R}+3e^{4\g+R/\rc}-13e^{R/\rc}R
-4(e^{4\g R}-1)(2e^{2\g R}-e^{R/\rc})\rc\Bigr)+
$$
$$
+4\g^2
\Bigl(2R^2-(3e^{4\g R}-5)R\rc+3(e^{4\g R}-1)\rc^2\Bigr)+
$$
$$
+32\g^3\rc
\Bigl(R^2-R\rc+e^{-R/\rc}(e^{4\g R}-1)(2e^{R/\rc}-e^{2\g R})\rc^2)\Bigr)
\Bigr].
$$
In calculating $I_{42}$, we put the condition
\be\l{cond2}
6\g\rc < 1, 
\ee
which is necessary to prevent divergency at the endpoint 
$\rbo=\infty$. Collecting the above two integrals we obtain 
\be
I_4 \equiv\C'_e =-\frac{A\g^3\rc^2}{6R(1-4\g^2\rc^2)^4}
\left[
e^{-2\g R}
\Bigl(
-R(3+2\g R(3+2\g R))
\right.
\ee
$$
+12\g^2R\rc^2(5+2\g R(5+2\g R))
-48\g^4R\rc^4(15+2\g R(7+2\g R))
$$
$$
\left.
+64\g^5\rc^6(24+\g R(33+2\g R(9+2\g R)))\Bigr)
-1536\g^5\rc^6 e^{-R/\rc}
\right].
$$
Thus, we have finally for the Coloumb integral for
exponential screened Coloumb potential,
\be\l{Coloumb4}
\C'_e =
-\frac{A\lambda^2}{8(1-\lambda^2)^4}
\frac{\g e^{-2\rho}}{\rho}
\left[
-(\rho+2\rho^2+\frac{4}{3}\rho^3)
+3\lambda^2(5\rho+10\rho^2+4\rho^3)
\right.
\ee
$$
-\lambda^4(15\rho+14\rho^2+4\rho^3)
\left.
+\lambda^6(8+11\rho+6\rho^2+\frac{4}{3}\rho^3 
- 8e^{2\rho-\frac{2\rho}{\lambda}})
\right].
$$
Here, we have used notation $\lambda=2\g r_c$, and 
also $\lambda<1/3$ due to Eq.(\ref{cond2}).
%
\begin{figure}
\plot{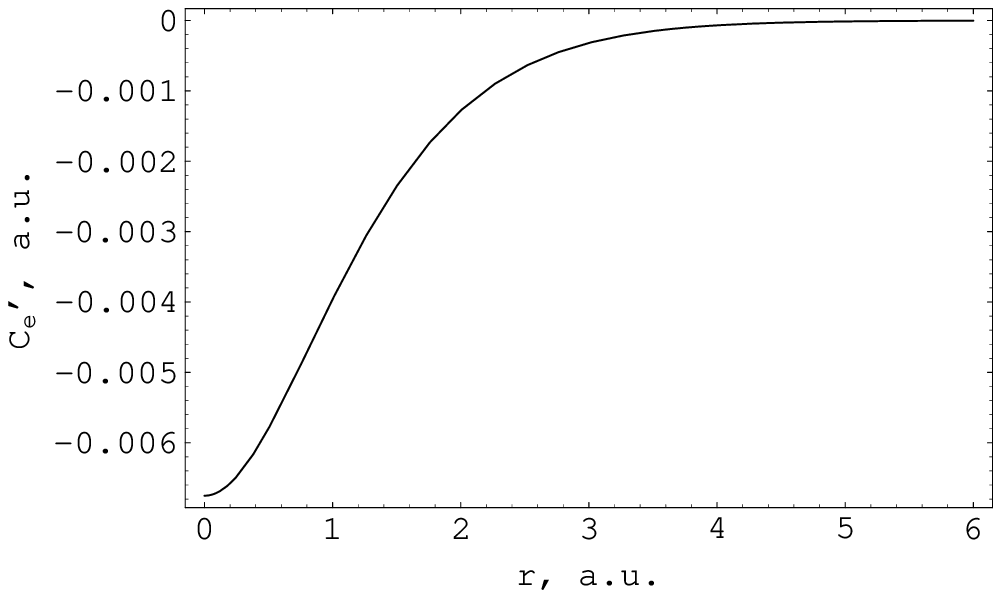}
\caption{The Coloumb integral $\C'_e$ as a function of $\rho$, 
Eq. (\ref{Coloumb4}), at $\lambda=1/37$. 
Here, $\rho=\gamma R$, where $R$ is the internuclear distance,
and $\lambda=2\gamma r_c$, where $r_c$ is the correlation length parameter.}
\label{Fig5}
\end{figure}
%
%
\begin{figure}
\plot{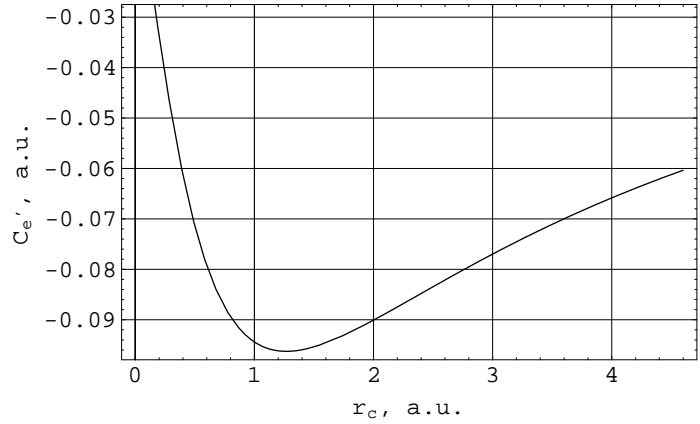}
\caption{The Coloumb integral $\C'_e$ as a function of $r_c$, 
Eq. (\ref{Coloumb4}), at $\rho=1.67$. For $r_c>0.2 \au$, the regularized
values of $\C'_e$ are presented.}
\label{Fig6}
\end{figure}
\clearpage
%
\begin{figure}
\plot{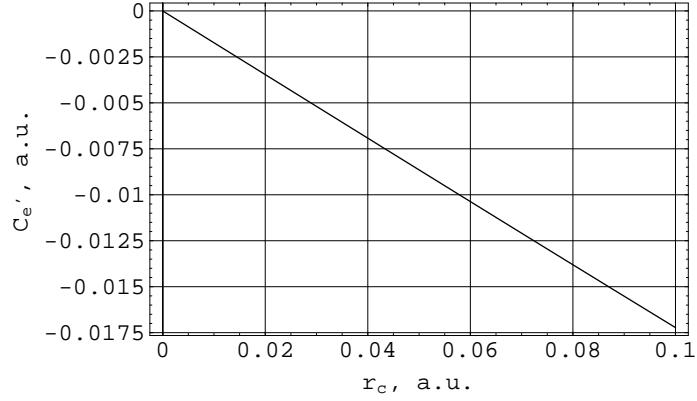}
\caption{The Coloumb integral $\C'_e$ as a function of $r_c$, 
Eq. (\ref{Coloumb4}), at $\rho=1.67$. More detailed view.}
\label{Fig7}
\end{figure}
The total Coloumb integral is
\be
\C'_{E} = \C'_C-\C'_e,
\ee
where $\C'_C$ is wellknown Coloumb potential part given by Eq.(\ref{C'0}).

Below, we turn to the other potential, Gaussian screened Coloumb potential,
considered by Santilli and Shillady \cite{SS}. The Coloumb integral 
for this potential can be calculated exactly, and the result contains 
one special function, the error function $\erf(z)$.

\subsubsection{Coloumb integral for Gaussian screened Coloumb potential}
\l{ColoumbGauss}

In this Section, we calculate the Coloumb integral
for the case of Gaussian screened potential.
Namely, we approximate the general potential (\ref{V12}) by \cite{SS}
\be\l{VGauss}
V(r_{12}) = V_C+V_g=\frac{e^2}{r_{12}}-\frac{Ae^{-r^2_{12}/c}}{r_{12}},
\ee
where $A$ and $c=\rc^2$ are positive parameters.
At long distances, $r_{12}\to\infty$, we can ignore $V_g$ while 
its series expansion about the point $r_{12}=0$ is
\be
V(r_{12})_{|r_{12}\to 0} = 
\frac{e^2-A}{r_{12}}+\frac{A}{c}r_{12}+ O(r^2_{12}).
\ee
Here, we should put $A=V_0r_c$ to have
the same coefficient at $r^{-1}_{12}$ in the $r_{12}\to 0$ asymptotics
as it is in the case of Hulten potential; see Eq.(\ref{series}).

The Coloumb integral is
\be\l{ColoumbG}
\C'_{G} = \int dv_1dv_2\ 
\left(\frac{e^2}{r_{12}}-\frac{Ae^{-r^2_{12}/c}}{r_{12}}\right)
|f(r_{a1})|^2|f(r_{b2})|^2.
\ee
The integral we are calculating is
\be
\C'_{g}=\int dv_1dv_2\ \frac{Ae^{-r^2_{12}/c}}{\r} |f(\rao)|^2|f(\rbt)|^2,
\ee
where notation and coordinate system are due to Sec.~\ref{ColoumbExp}.
First, we integrate over angular coordinates of second electron,
\be
I_1 = \int\limits_{0}^{2\pi}d\pt\int\limits_{0}^{\pi}d\tt \ 
\frac{Ae^{-r^2_{12}/c}}{\r}\frac{\g^3}{\pi}e^{-2\g\rbt}\rbt^2\sin\tt
,
\ee
where $\r$ is defined by (\ref{r12app}). We have
\be
I_1 = \frac{A\g^3\sqrt{\pi c}e^{-2\g\rbt}}{\rbo}
\left(
\erf(\sqrt{\frac{(\rbo+\rbt)^2}{c}})-
\erf(\sqrt{\frac{(\rbo-\rbt)^2}{c}})
\right),
\ee
where 
\be
\erf(z) = \frac{2}{\sqrt{\pi}}\int\limits_{0}^{z}e^{-t^2}dt
\ee
is error function.
Further, integrating on radial coordinate $\rbt$ must be 
performed in separate intervals,
\be
I_2 =\int\limits_{0}^{\rbo}d\rbt\ I_1(\rbt<\rbo)+
\int\limits_{\rbo}^{\infty}d\rbt\ I_1(\rbt>\rbo),
\ee
where 
\be
\sqrt{(\rbo-\rbt)^2} = 
\left\{ 
\begin{array}{cc}
r_{b1}-r_{b2}, & \quad r_{b2}<r_{b1},\\
r_{b2}-r_{b1}, & \quad r_{b2}>r_{b1},
\end{array}
\right.
\ee
with the result
\be
I_2= -\frac{A\g\sqrt{c}e^{-2\g\rbo-\rbo^2/c}}{4\rbo}
\Biggl(
4\g\sqrt{c}(e^{\rbo^2/c}-e^{2\g\rbo)})
\ee
$$
+\sqrt{\pi}e^{\rbo^2/c+c\g^2}
\Biggl[(1+2\g(\rbo-c\g))(\erfc(\frac{\rbo-c\g}{\sqrt{c}})+2\erfc(\sqrt{c}\g)-2)
$$
$$
+e^{4\g\rbo}(2\g(\rbo+c\g)-1)\erfc(\frac{\rbo+c\g}{\sqrt{c}})
\Biggr]
\Biggr),
$$
where $\erfc(z)=1-\erf(z)$.
Now, we turn to integrating over coordinates of first electron,
$(\rbo, \tone, \po)$,
\be
I_3 =  \int\limits_{0}^{\pi}d\tone\int\limits_{0}^{2\pi} d\po\ 
I_2\frac{\g^3}{\pi}e^{-2\g\rao}\rbo^2\sin\tone,
\ee
where $\rao$ is defined by (\ref{ra1app}). We obtain
after tedious calculations
\be
I_3=
-\frac{A\sqrt{c}\g^2}{8R}
e^{-\frac{\rbo^2}{c}-2\g(\sqrt{(R-\rbo)^2}+2\rbo+\sqrt{(R+\rbo)^2})}
\ee
$$
\times\Biggl(
e^{2\g(\sqrt{(R-\rbo)^2}+\rbo)}
-e^{2\g(\sqrt{(R+\rbo)^2}+\rbo)}
-2\g e^{2\g(\sqrt{(R+\rbo)^2}+\rbo)}\sqrt{(R-\rbo)^2}
$$
$$
+2\g e^{2\g(\sqrt{(R-\rbo)^2}+\rbo)}\sqrt{(R+\rbo)^2}
\Biggr)
$$
$$
\times\Biggl[
\sqrt{\pi}e^{\frac{\rbo^2}{c}+c\g^2}
\Biggl(
(1+2\g\rbo-2c\g^2)(2\erfc(\g\sqrt{c})+\erfc(\frac{\rbo-\g c}{\sqrt{c}})-2)
$$
$$
+(1+2\g\rbo+2c\g^2)e^{4\g\rbo}\erfc(\frac{\rbo+\g c}{\sqrt{c}})
\Biggr)
\Biggr].
$$
Again, we must further integrate in $\rbo$ by separate intervals,
\be\l{I4G}
I_4 =\int\limits_{0}^{R}d\rbo\ I_3(\rbo<R)+
\int\limits_{R}^{\infty}d\rbo\ I_3(\rbo>R).
\ee
First, we replace the endpoint $\rbo=\infty$ by finite value
$\rbo=\Lambda$ to avoid divergencies at intermediate calculations.
After straightforward but tedious calculations
we obtain rather long expression so that we do not represent it
here noting however that the following integrals are used
during the calculations:
\be
\int \erf(z)dz = \frac{e^{-z^2}}{\sqrt{\pi}} + z\ \erf(z),
\ee 
\be
\int z\ \erf(z)dz = \frac{ze^{-z^2}}{2\sqrt{\pi}} -\frac{1}{4}\erf(z)
+\frac{1}{2} z^2\erf(z),
\ee
\be
\int e^{-az}\erf(z)dz = -\frac{1}{a}e^{-az}\erf(z) 
+ \frac{1}{a}e^{a^2/4}\erf(\frac{a}{2}+z),
\ee
\be
\int z e^{-az}\erf(z)dz = -\frac{1}{a\sqrt{\pi}}e^{-az-z^2} 
-\frac{1}{a^2}e^{-az}(1+az)\erf(z)
\ee
$$
-\frac{1}{2a^2}(a^2-1)e^{a^2/4}\erf(\frac{a}{2}+z),
$$
\be
\int e^{-az-bz^2} dz =
\frac{\sqrt{\pi}}{2\sqrt{b}}e^{a^2/(4b)}\erf(\frac{a+2bz}{2\sqrt{b}}),
\ee
\be
\int z e^{-az-bz^2} dz =
-\frac{1}{2b}e^{-az-bz^2}-
\frac{a\sqrt{\pi}}{4b^{3/2}}e^{a^2/(4b)}\erf(\frac{a+2bz}{2\sqrt{b}}).
\ee
Using $\lim_{\Lambda\to\infty}\erf(\Lambda)=1$ and replacing
welldefined exponentially decreasing terms by zero,
we obtain some finite terms and big number (about fourty) of 
$\Lambda$ dependent terms, which 
are unbounded at $\Lambda\to\infty$. All the divergent terms
totally cancel each other so the final expression turns 
out to be automatically finite.
 
As the result, we obtain the Coloumb integral for
Gaussian screened Coloumb potential in the following form:
\be\l{ColoumbGa}
\C'_{g}=
\frac{A\g\k e^{-2\rho}}{96\rho}
\Biggl[
-(60+96\rho+48\rho^2)\k+(32+48\rho)\k^3-16\k^5
\ee
$$
+\Bigl((60+16\rho^2)\k - 32\k^3+16\k^5\Bigr)e^{2\rho-\frac{\rho^2}{\k^2}}
$$
$$
+\sqrt{\pi}e^{\k^2}
\Biggl(
(30\rho+8\rho^3-36\rho\k^2+24\rho\k^4)
(2\erf(\k)-\erfc(\frac{\rho}{\k}-\k)-e^{4\rho}\erfc(\frac{\rho}{\k}+\k))
$$
$$
+(15+24\rho^2-(18+24\rho^2)\k^2+12\k^4-8\k^6)
(2\erf(\k)-\erfc(\frac{\rho}{\k}-\k)+e^{4\rho}\erfc(\frac{\rho}{\k}+\k))
\Biggr)
\Biggr],
$$
where we have used notation 
\be
\k=\g\sqrt{c}= \g\rc=\frac{\lambda}{2}.
\ee
%
\begin{figure}
\plot{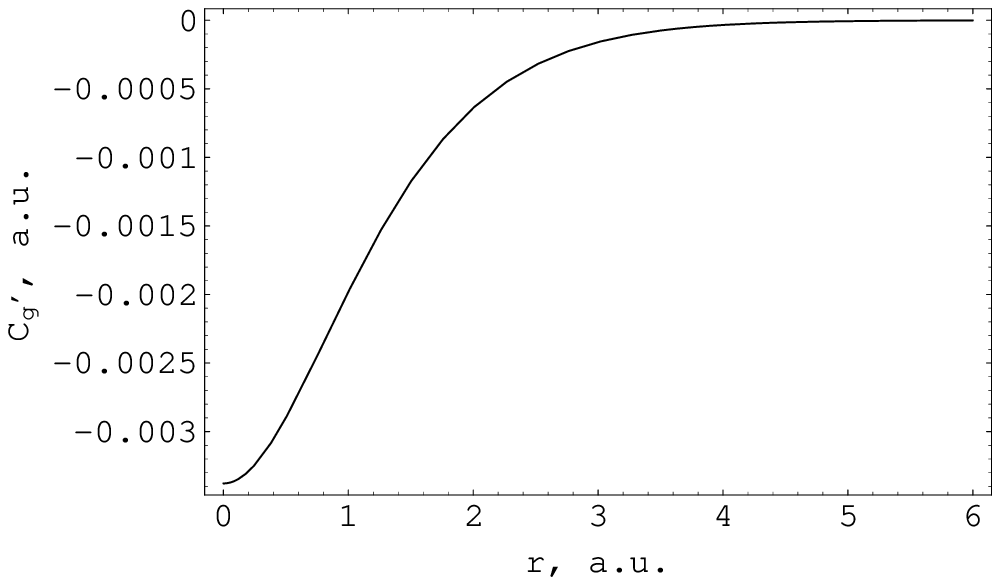}
\caption{The Coloumb integral $\C'_g$ as a function of $\rho$, 
Eq. (\ref{ColoumbGa}), at $2\kappa=\lambda=1/37$. 
Here, $\rho=\gamma R$, where $R$ is the internuclear distance,
and $\lambda=2\gamma r_c$, where $r_c$ is the correlation length parameter.}
\label{Fig8}
\end{figure}
%
%
\begin{figure}
\plot{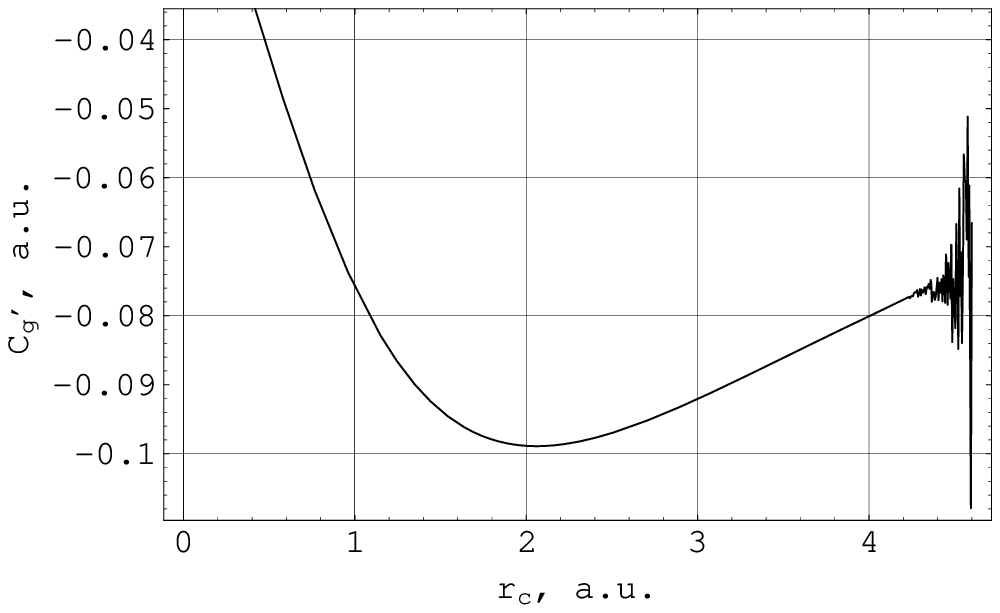}
\caption{The Coloumb integral $\C'_g$ as a function of $r_c$, 
Eq. (\ref{ColoumbGa}), at $\rho=1.67$.}
\label{Fig9}
\end{figure}
%
%
\begin{figure}
\plot{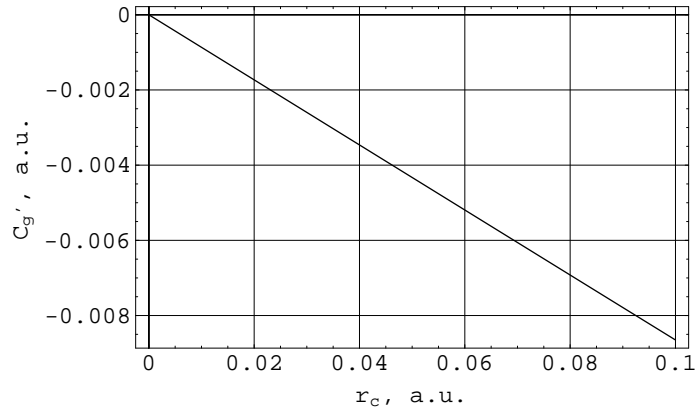}
\caption{The Coloumb integral $\C'_g$ as a function of $r_c$, 
Eq. (\ref{ColoumbGa}), at $\rho=1.67$. More detailed view.}
\label{Fig10}
\end{figure}

The total Coloumb integral is 
\be
\C'_G = \C'_C-\C'_g,
\ee
where $\C'_C$ is given by Eq.(\ref{C'0}).

\clearpage

\subsubsection{Exchange integral}
\l{Exchange}

Our general remark is that all calculations for the above 
{\it Coloumb} integrals
are made in spherical coordinates, which correspond to spherical symmetry
of the charge distributions of both $1s$ electrons, $|\psi(r_{a1})|^2$
and $|\psi(r_{a2})|^2$, each moving around one nucleus.  
One can use prolate spheroidal coordinates, which are exploited sometimes 
when integrating over coordinates of last electron, but we have encountered 
the same problem of big number of terms in the intermediate expressions,
with no advantage in comparison to the use of spherical coordinates.

Unlike to Coloumb integral, calculation of {\it exchange} integral should 
be made in the spheroidal coordinates, which correspond to spheroidal 
symmetry of charge distributions of the electrons,
$\psi^*(r_{a1})\psi(r_{b1})$ and $\psi^*(r_{a2})\psi(r_{b2})$, each moving
around {\it both} the nuclei, $a$ and $b$.

Calculation of the exchange integral, 
\be\l{exchange3}
\E' = \int dv_1dv_2\ V(r_{12})f^*(r_{a1})f(r_{b1})f^*(r_{a2})f(r_{b2}),
\ee
essentially depends on the form of the potential $V(r_{12})$
in the sense that the integration can be made only in {\it spheroidal}
coordinates, $(x_1,y_1,\varphi_1)$ and $(x_2,y_2,\varphi_2)$,
and one should use an expansion of $V(r_{12})$ in the 
associated Legendre polynomials.

For the usual Coloumb potential, $V(r_{12})=r^{-1}_{12}$, 
it is rather long (about 12 pages to present the main details) 
and nontrivial calculation, where Neumann expansion in terms of 
associated Legendre polynomials, in spheroidal coordinates, is used 
(celebrated result by Sugiura, see Eq.(\ref{Sugiura})).

In general, any analytical square integrable function can be 
expanded in associated Legendre polynomials. However, in direct
calculating of the expansion coefficients by means of integral
of the function with Legendre polynomials, one meets serious 
problems even for simple functions. Practically, one uses, instead,
properties of special functions to derive such expansions.

We mention that there is Gegenbauer expansion \cite{Varshalovich},
having in a particular case the form \cite{Flugge}
\be\l{Gegenbauer}
\frac{e^{ikr_{12}}}{r_{12}}
=\frac{1}{r_1r_2}\sum\limits_{l=0}^{\infty}
\sqrt{\frac{2l+1}{4\pi}}\frac{i}{k}j_l(kr_1)n^{(1)}_l(kr_2)
Y_{l,0}(\theta_{12}),
\ee
where $j_l(z)$ and $n^{(1)}_{l}(z)$ are spherical Bessel
and spherical Hankel functions of first kind, respectively,
$\theta_{12}$ is angle between vectors $\vec r_1$ and $\vec r_2$,
and $r_1=|\vec r_1|$, $r_2=|\vec r_2|$; $r_1<r_2$.
Spherical harmonics $Y_{l,0}(\theta_{12})$ can be rewritten
in terms of Legendre polynomials due to the summation theorem.

We note that this expansion can be used, at $k=i/r_c$, 
to reproduce exponential screened potential, $V_e(r_{12})$,
and to calculate associated exchange integral (\ref{exchange3})
but, however, it concerns {\it spherical} (not spheroidal) coordinates,
$(r_1,\theta_1,\varphi_1)$ and $(r_2,\theta_2,\varphi_2)$.

For Hulten potential $V_h(r_{12})$, exponential screened potential, 
$V_e(r_{12})$, and Gaussian screened potential, $V_g(r_{12})$, 
which are of interest in this paper, we have no such an expansion 
in {\it spheroidal} coordinates. To stress that this is not only 
the problem of changing coordinate system, we mention that the 
solution of usual 3-dimensional wave equation, $\Delta\psi+k^2\psi=0$,
is given by function $e^{i\vec k\vec r}/r$, in spherical coordinates, 
to which one can apply Gegenbauer expansion, 
while in spheroidal coordinates its solution is represented
by complicated function containing infinite series 
of recurrent coefficients \cite{Abramovits};
see also \cite{ibr6}, Sec.~3.4. As the result, we have no
possibility to calculate exactly exchange integrals for 
these non-Coloumb potentials.

In order to obtain {\it approximate} expression for the exchange integral 
for the case of the above non-Coloumb potentials, we make analysis 
of asymptotics of the standard exchange integral (i.e. that for the 
Coloumb potential), Eq.(\ref{Sugiura}). It is easy to derive that 
\be\l{rhoinfty}
{\E'_C}_{|\rho\to\infty} \sim e^{-2\rho}, 
\ee
at long distances between the nuclei, 
and 
\be\l{5to8}
{\E'_C}_{|\rho=0} = \frac{5}{8}\g,
\ee
in the case of coinciding nuclei.
At $r^{-1}_c\to 0$, we should have the same asymptotics for 
exchange integral for each of the above non-Coloumb potentials because
these potentials behave as Coloumb potential at $r^{-1}_c\to 0$.

In both the limiting cases, $\rho\to\infty$ and $\rho=0$,
the exchange integral for the non-Coloumb potentials is simplified, 
and one can use spherical coordinates since the two-center problem
is reduced to one-center problem. We consider two limiting cases.

{\bf a) $\rho=\infty$ case.}

This case is trivial because exchange integral tends to zero
due to lack of overlapping of the wave functions of two $H$ atoms. 

{\bf b) $\rho=0$ case.}

In this case , we have $r_{a1}=r_{b1}=r_1$ and 
$r_{a2}=r_{b2}=r_2$ so that Eq.(\ref{exchange3}) becomes
\be\l{exchange4}
\E' = \int dv_1dv_2\ V(r_{12})|f(r_1)|^2|f(r_2)|^2,
\ee
One can see that this is the case of $He$ atom with two electrons 
in the ground state.
Evidently, in terms of our anzatz (\ref{wf}) we have complete overlapping 
of the wave functions.

Even the above mentioned simplification of the exchange integral and 
use of spherical coordinates does not enable us to calculate 
{\it straightforwardly} the integral (\ref{exchange4}) for the non-Coloumb 
potentials, $V_h$, $V_e$, or $V_g$; the integrands are still too complicated. 
This indicates that we should use expansion of these potentials in 
Legendre polynomials, in spherical coordinates, to perform the integrals. 
Only exponential screened potential $V_e$ is given such an expansion 
here.
Namely, this is Gegenbauer expansion (\ref{Gegenbauer}), owing to which
we can calculate the exchange potential for exponential screened 
potential $V_e$, to which we turn below.\\

{\bf Exchange integral for the exponential screened Coloumb potential
$V_e$, at $\rho=0$.}\\

The integral is 
\be\l{exchange5}
{\E'_E}_{|\rho=0} \equiv
(\E'_C-\E'_e)_{|\rho=0}
=\frac{5}{8}\g - 
\int dv_1dv_2\ \frac{Ae^{-r_{12}/r_c}}{r_{12}}|f(r_1)|^2|f(r_2)|^2,
\ee
where we have used Eq.(\ref{5to8}) for the usual Coloumb potential part 
of the integral.
In the Gegenbauer expansion (\ref{Gegenbauer}), we assume $k=i/r_c$
to reproduce the potential $V_e(r_{12})$.
Since the wave functions $f(r_1)$ and $f(r_2)$ given by 
Eq.(\ref{wf0}) do not depend on the angles, only $l=0$, $m=0$ term
of the expansion (\ref{Gegenbauer}) contributes to the exchange integral
(\ref{exchange5}) due to orthogonality of Legendre polynomials.
Using 
\be
j_0(z)= \sin z, \quad n_0(z)=-ie^{iz}, \quad Y_{0,0}= \sqrt{\frac{1}{4\pi}},
\ee
we thus have 
\be
\frac{Ae^{ikr_{12}}}{r_{12}} \to 
\left\{ 
\begin{array}{cc}
\frac{A}{kr_1r_2} \sin kr_1 e^{ikr_2} , & \quad r_1<r_2,\\
\frac{A}{kr_1r_2} \sin kr_2 e^{ikr_1} , & \quad r_1>r_2,
\end{array}
\right.
\ee
Then the exchange integral (\ref{exchange5}) is written as
\be
{\E'_E}_{|\rho=0} = \frac{5}{8}\g - 
\int\limits_{0}^{\infty} 4\pi r^2_2 dr_2 \Bigl[
\int\limits_{0}^{r_2} 4\pi r^2_1 dr_1 
\frac{A}{kr_1r_2} \sin kr_1 e^{ikr_2} 
\frac{\g^3}{\pi}e^{-2\g r_1}
\frac{\g^3}{\pi}e^{-2\g r_2}
\ee
$$
+\int\limits_{r_2}^{\infty} 4\pi r^2_1 dr_1 
\frac{A}{kr_1r_2} \sin kr_2 e^{ikr_1}
\frac{\g^3}{\pi}e^{-2\g r_1}
\frac{\g^3}{\pi}e^{-2\g r_2}\Bigr],
$$
where $4\pi r^2_1$ and $4\pi r^2_2$ are volume factors.
The two above integrals over $r_1$ can be easily calculated,
with the result
\be
\frac{16A\g^6r_2}{(k^2+4\g^2)^2}
\Bigl[
4\g e^{i(k+2i\g)r_2}
+\frac{1}{k}e^{i(k+4i\g)}
\Bigl(
(k^2-4\g^2)\sin kr_2 - 4k\g\cos kr_2 
\ee
$$
- (k^2+4\g^2)(k\cos kr_2 + 2\g\sin kr_2)
\Bigr)\Bigr]
$$
and 
\be
-\frac{16A\g^6r_2}{k(k+2i\g)^2}
\Bigl(1+(2\g-ik)\sin kr_2 e^{i(k+4i\g)}\Bigr).
\ee
Summing up these terms and integrating over $r_2$ we get
after some algebra
\be
{\E'_E}_{|\rho=0} = \frac{5}{8}\g
+\frac{A\g^3}{2(k+2i\g)^4}(k^2+8ik\g-20\g^2).
\ee
Inserting 
\be
k=\frac{i}{r_c},
\ee
to reproduce the potential $V_e$,
and denoting $\lambda = 2\g r_c$ we write down our final result,
\be\l{EE0}
{\E'_E}_{|\rho=0} = \frac{5}{8}\g 
- \frac{\g A\lambda^2}{8(1+\lambda)^4}(1+4\lambda+5\lambda^2).
\ee
Note that, at $r^{-1}_c\to 0$, i.e. at $\lambda\to\infty$, we have 
\be\l{E5to8}
{\E'_E}_{|\rho=0}=\frac{5}{8}\g -\frac{5}{8}A\g
\ee
that is in agreement with the value (\ref{5to8}). 
We should to emphasize here that Eq.(\ref{EE0}) is {\it exact} result 
for the exchange integral $\E'_E$, at $\rho=0$. 

Next step is to implement $\rho$ dependence into (\ref{EE0}) following 
to natural criteria.
To restore partially $\rho$ dependence in the exchange integral (\ref{EE0}),
we use exact result (\ref{Sugiura}), 
and write down for the $\rho$ dependent exchange integral the following
{\it approximate} expression:
\be\l{EE}
\E'_E = \E'_C-\E'_e \approx
\E'_C -\frac{A\lambda^2}{(1+\lambda)^4}
(\frac{1}{8}+\frac{1}{2}\lambda+\frac{5}{8}\lambda^2)\frac{8}{5}\E'_C,
\ee
where $\E'_C$ is standard exact exchange integral for Coloumb potential
given by Eq.(\ref{Sugiura}) while the approximate $\lambda$ dependent part
arised from our potential $V_e$.
%
\begin{figure}
\plot{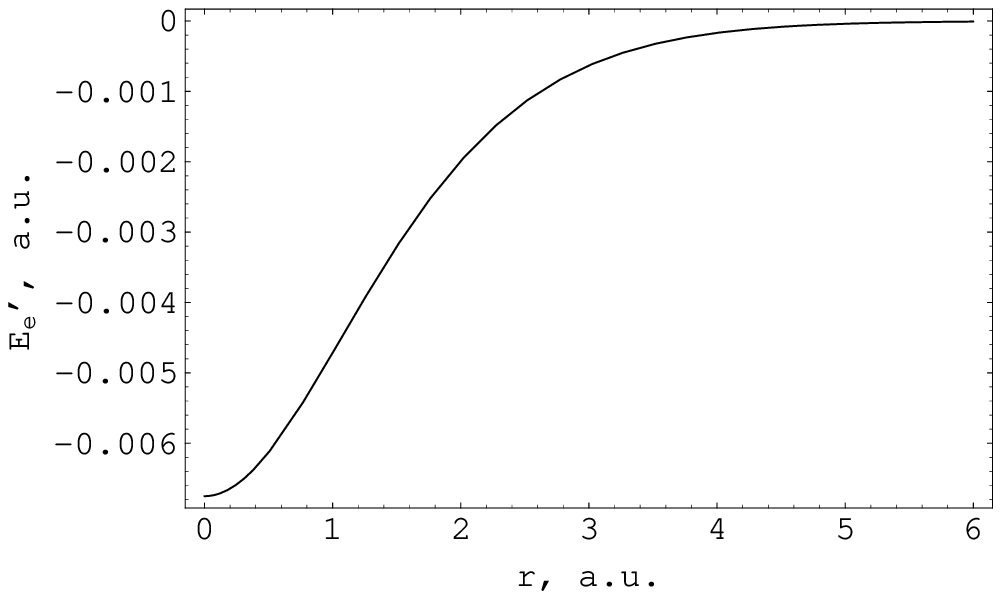}
\caption{The exchange integral $\E'_e$ as a function of $\rho$, 
Eq. (\ref{EE}), at $\lambda=1/37$. 
Here, $\rho=\gamma R$, where $R$ is the internuclear distance,
and $\lambda=2\gamma r_c$, where $r_c$ is the correlation length parameter.}
\label{Fig11}
\end{figure}
%
%
\begin{figure}
\plot{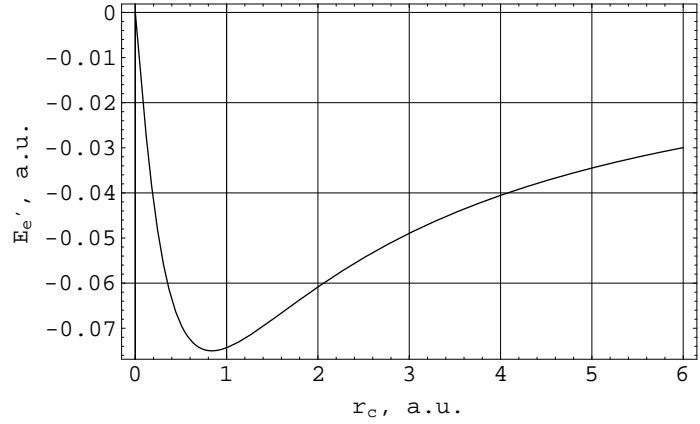}
\caption{The exchange integral $\E'_e$ as a function of $r_c$, 
Eq. (\ref{EE}), at $\rho=1.67$.}
\label{Fig12}
\end{figure}
%
%
\begin{figure}
\plot{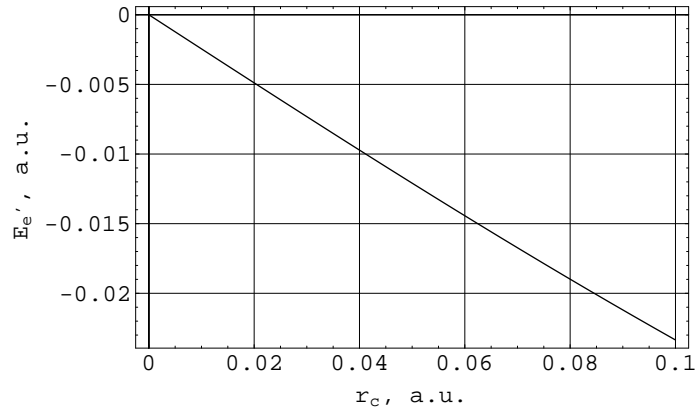}
\caption{The exchange integral $\E'_e$ as a function of $r_c$, 
Eq. (\ref{EE}), at $\rho=1.67$. More detailed view.}
\label{Fig13}
\end{figure}
\clearpage

We have a good accuracy of the approximation (\ref{EE}). Indeed, 
exchange integrals make sensible contribution to the total 
molecular energy at deep overlapping of the wave functions, $S>0.5$,
and we have made calculation just for the case of complete 
overlapping, $S=1$, with the necessary asymptotic factor, $e^{-2\rho}$,
provided by $\E'_C(\rho)$. Note that at $\lambda\to\infty$, the term
$\E'_e$ of Eq.(\ref{EE}) becomes $A\E'_C$, as it should be because
at $\lambda\to\infty$ (no screening) we have $V_e \to A/r_{12}$.
 In addition, although there is no possibility
to restore completely $\rho$ dependence for the second term in r.h.s.
of Eq.(\ref{EE}), we have got information on $\lambda$ dependence, 
which is of {\it most} interest here.

\subsection{Numerical calculations for the $V_e$-based model}
\l{Numerical}

In this Section, we consider the case of exponential screened 
potential $V_e=Ae^{-r_{12}/r_c}/r$,
for which we have calculated all the needed molecular integrals.

The $H_2$ molecule energy, due to Eq.(\ref{Emol}),
is written as
\be\l{EmolE}
E_{mol}(\g,R,A,r_c)
= 2\frac{\A+\A'\S}{1+\S^2} 
- \frac{2(\C+\E\S)-(\C'_C-\C'_e+\E'_C-\E'_e)}{1+\S^2}
+\frac{1}{R},
\ee
where the specific terms are the Coloumb integral $\C'_e$ given by 
Eq.(\ref{Coloumb4}) and the exchange integral $\E'_e$ given by (\ref{EE}).
We should find extremum of $E_{mol}$ as a function of our
basic parameters, $\g$, $R$, $A$, and $r_c$. We are using notation
$\rho=\g R$ and $\lambda = 2\g r_c$ so that our four parameters are
$\g$, $\rho$, $A$, and $\lambda$. In general, 
the number of energy levels of isoelectronium can also be viewed
as a parameter of the model. However, we restrict our consideration
by the one-level case, $\beta^2=1$; see Sec.~\ref{SecA}.

\subsubsection{Minimization of the energy}

First, we analyze the $A$ dependence of $E_{mol}$.
Due to Eq.(\ref{Arc}), for {\it one-level} isoelectronium we have 
$A=r_c^{-1}$, that can be identically rewritten as
\be\l{Alambda}
A=\frac{2\g}{\lambda}.
\ee
Thus the $A$ dependence converts to $\g$ and $\lambda$ dependence. 
This is the consequence of consideration of the Hulten potential
interaction for the electron pair made in Sec.~\ref{SecA}.

Second, we turn to $\gamma$ dependence. Due to (\ref{Alambda}),
the $A$ dependent parts, $\C'_e$ and $\E'_e$, acquire additional 
$\g$ factor and thus become $\g^2$ dependent. The other molecular 
integrals depend on $\g$ linearly so that we define accordingly,
\be
\bar\C = \frac{1}{\g}\C, \quad
\bar\E = \frac{1}{\g}\E, \quad
{\bar\C}'_C = \frac{1}{\g}\C'_C, \quad
{\bar\E}'_C = \frac{1}{\g}\E'_C, \quad
{\bar\C}'_e = \frac{1}{\g^2}\C'_e, \quad
{\bar\E}'_e = \frac{1}{\g^2}\E'_e.
\ee
Inserting the computed integrals $\A$ and $\A'$ 
into (\ref{EmolE}) we have \be\l{EmolE2}
E_{mol}(\g,\rho,\lambda) = -a\g+ b\g^2,
\ee 
where 
\be
a(\rho,\lambda) 
= \frac{2+2{\bar\C}+4S{\bar\E}-{\bar\C}'_C-{\bar\E}'_C}{1+S^2} 
- \frac{1}{\rho}
\ee
and 
\be
b(\rho,\lambda) 
= \frac{S^2-1 -2S{\bar\E}+{\bar\C}'_e+{\bar\E}'_e}{1+S^2}. 
\ee
The value of $\g$ corresponding to an extremum of $E_{mol}$ is found 
from the equation $dE_{mol}/d\g=0$, which gives the optimal value
\be
\g_{opt} = \frac{a}{2b}.
\ee
Inserting this into (\ref{EmolE2}) we get the extremal
value of $E_{mol}$,
\be
E_{mol}(\rho,\lambda) = -\frac{a^2}{4b}.
\ee
Using definitions of $a$ and $b$ we have explicitly
\be\l{g}
\g_{opt}= 
\frac{1-2\rho+S^2+\rho(-2{\bar\C}-4S{\bar\E}+{\bar\C}'_C+{\bar\C}'_C)}
{2\rho(-1+S^2-2S{\bar\E}'_C+{\bar\C}'_e+{\bar\E}'_e)}
\ee
and
\be\l{EmolNum}
E_{mol}(\rho,\lambda)= 
\frac{(1-2\rho+S^2+\rho(-2{\bar\C}-4S{\bar\E}+{\bar\C}'_C+{\bar\C}'_C))^2}
{4\rho^2(1+S^2)(-1+S^2-2S{\bar\E}'_C+{\bar\C}'_e+{\bar\E}'_e)}.
\ee
%
\begin{figure}
\plot{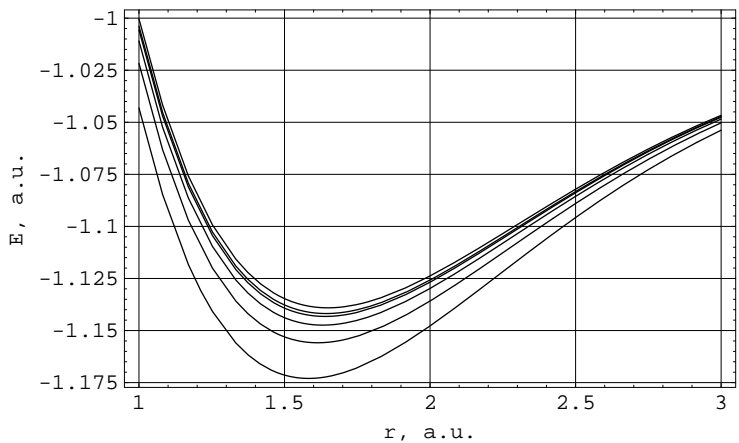}
\caption{The total energy $E=E_{mol}$ as a function of $\rho$, 
Eq. (\ref{EmolNum}), at $\lambda=1/60, 1/40, 1/20, 1/10, 1/5$. 
The lowest plot corresponds to $\lambda=1/5$ 
($\rho=\gamma R$, $\lambda=2\gamma r_c$).}
\label{Fig14}
\end{figure}

Next, we turn to the extremum in the parameter $\rho$. The $\rho$ dependence,
as well as the $\lambda$ dependence, of $E_{mol}$ is essentially 
nonalgebraic so that we are forced to use numerical calculations. 

It appears that the $\lambda$ dependence does not reveal any local 
energy minimum while the $\rho$ dependence does. 
Below, we use the condition, 
$\lambda^{-1}=$ {\it integer number}, obtained during the calculation
of the Coloumb integral with Hulten potential $V_h$; see Eq.(\ref{k}).
Although there is obviously no necessity to keep this condition 
for the case of exponential screened potential $V_e$, we consider it
as a prescription for allowed values of $\lambda$.

Since the $\lambda$ dependence of the energy has no minimum
we can use fitting of the predicted energy $E_{mol}(\lambda)$ 
to the experimental value by varying $\lambda$. 
This allows us to estimate the value of the parameter $\lambda$, and 
thus the value of the effective radius of the isoelectronium 
$r_c=\lambda/2\gamma_{opt}$.

\subsubsection{Fitting of the energy and the bond length}

The procedure is the following. We fix some numerical value
of $\lambda$, and identify minimal value of $E_{mol}(\rho,\lambda)$,
given by Eq.(\ref{EmolNum}), in respect with the parameter $\rho$. 
This gives us minimal energy and 
corresponding optimal value of $\rho$, at some fixed value of $\lambda$.
Then, we calculate $\g_{opt}$ by using Eq.(\ref{g}),
and use obtained values of $\rho_{opt}$ and $\g_{opt}$ 
to calculate values of $R_{opt}$ and $r_c$. 

We calculated minimal values of $E_{mol}$ in $\rho$,
for a wide range of integer values of $\lambda^{-1}$.  
The results are presented in Tables~2 and 3, and Figures~15 and 16. 
One can see that the energy $E_{mol}$ decreases with the increase of 
$r_c$ (proportional to size of isoelectronium), as it was expected to be. 

We note that all the presented values of $E_{mol}$ in Tables~2 and 3
are lower than that, $E_{mol}^{var}=-1.139 \au$, obtained via two-parametric 
Ritz variational approach  to the standard model of $H_2$ 
(see, e.g., \cite{Flugge}), which is the model without 
the assumption of short-range attractive potential between 
the electrons. This means that the $V_e$-based model gives 
better prediction than the one of the standard model,
for any admitted value of the effective radius of isoelectronium 
$r_c>0$. Indeed, the standard prediction $E_{mol}^{var}=-1.139 \au$ 
is much higher than the experimental value 
$E_{exper}[H_2]=-1.174474 \au$

\subsubsection{The results of fitting}

{\bf Best fit of the energy $E_{mol}$}.\\

Due to Table~2 (see also Fig.~15), the experimental value, 
$E_{exp}[H_2]=-1.174 ... -1.164 \au$ (here we take 0.9\% uncertainty
of the experimental value) is fitted by 
\be
r_c=0.0833 ... 0.0600 \au,
\ee
i.e. $\lambda= 1/5...1/7$, 
with the optimal distance, $R_{opt}=1.3184 ...1.3441 \au$
We see that the predicted $R_{opt}$ appeared to be about 6\% less 
than the experimental value $R_{exper}[H_2]=1.4011 \au$
We assign this discrepancy to the approximation we have made
for the exchange integral (\ref{EE}).

Below we fit $R_{opt}$, to estimate the associated minimal energy.\\

{\bf Best fit of the internuclear distance $R$}.\\

Due to Table~2 (see also Fig.~16), the experimental value of the 
internuclear distance, $R_{exp}=1.4011 \au$, is fitted by 
$r_c=0.0115 \au$, 
with the corresponding minimal energy $E_{min}=-1.144 \au$,
which is about 3\% bigger than the experimental value. 
Again, we assign this discrepancy to the approximation we have made 
for the exchange integral (\ref{EE}), and take 
\be 
r_c=0.0115 \au, 
\ee
i.e. $\lambda=1/37$, as the result of our final fit noting
that (a) in ref. \cite{SS} the value $r_c=0.0112 \au$ has been used to make 
exact numerical fit of the energy, with corresponding $R =1.40 \au$,
and (b) we have less discrepancy.\\

{\bf The weight of the pure isoelectronium phase.}\\

To estimate the weight of the pure isoelectronium phase, 
which can be viewed as a measure of stability of 
the pure isoelectronium state, 
we use the above obtained fits and the fact that this phase makes 
contribution to the total molecular energy via the Coloumb and 
exchange integrals. 

According to Eq.(\ref{EmolE}), the isoelectronium phase displays itself
only by the term $P_e\equiv |\C'_e(\g,\rho,\lambda)+\E'_e(\g,\rho,\lambda)|$ 
while the Coloumb phase displays itself by the corresponding term 
$P_C\equiv |\C'_C(\g,\rho)+\E'_C(\g,\rho)|$.
Putting the total sum $P_C+P_e=1$, i.e. $P_C+P_e$ is 100\%, 
the weights are defined simply by
\be
W_C = \frac{P_C}{P_C+P_e}, \quad W_e = \frac{P_e}{P_C+P_e},
\ee

{\it The weight for the best fit of $R$.}\\
At the values $\lambda=1/37$ (i.e. $\rc=0.0115$), $\g=1.1706$, and 
$\rho=1.6320$, for which we have minimal $E_{mol}=-1.144$ and 
optimal $R=1.40$, we get the numerical values of the weights, 
\be
W_e=0.84\% 
\ee
for the pure isoelectronium phase, 
and $W_C=100\%-W_e=99.16\%$ for the Coloumb phase.

{\it The weight for the best fit of $E_{mol}$.}\\
At the values $\lambda=1/5$ (i.e. $\rc=0.0833 \au$), $\g=1.2005$, and 
$\rho=1.5827$, for which we have minimal $E_{mol}=-1.173 \au$ and 
optimal $R=1.318 \au$, we obtain 
\be
W_e=6.16\%, \quad W_C=93.84\%.
\ee
From the above two cases, one can see that the weight of pure 
isoelectronium phase is estimated to be 
\be
W_e \simeq 1...6\%,
\ee
for the predicted variational energy $E_{mol}=-1.143...-1.173 \au$

{\it The biggest possible weight.}\\
Note that in our $V_e$-based model the biggest allowed value 
of $\lambda$ is $\lambda=1/4$
(i.e. $\rc=0.1034$) because $\lambda<1/3$, to avoid divergency
of the Coloumb integral $\C_e$. 
For this value of $\lambda$, we obtain minimal 
$E_{mol}=-1.182 \au$ and optimal $R=1.297 \au$
This value corresponds to the {\it biggest} possible weight of 
the pure isoelectronium phase, 
\be
W_e=7.32\%, 
\ee
within our approximate model. 

The following three remarks are in order.

(i) We consider the existence of this upper limit, 
$W_e \leq 7.32\%$, as a highly remarkable 
implication of our $V_e$-model noting however that it may be artifact 
of the use of the exponential screened Coloumb potential.

(ii) Another remarkable implication is due to the condition,
$\lambda^{-1}=$ {\it integer number}, obtained for the case
of Hulten potential. One can see from Table~2 that the energy $E_{mol}$
varies {\it discretely} with the discrete variation of $\lambda^{-1}$.
This means that there is no possibility to make a ``smooth fit".
For example, at $\lambda=1/5$, we have $E_{mol}=-1.173$, 
and the {\it nearest} two values, $\lambda=1/4$ and $\lambda=1/6$,
give us $E_{mol}=-1.182$ and $E_{mol}=-1.167$, respectively.
Therefore, owing to the above condition the model becomes 
{\it more predicitive}.

(iii) Numerical calculation shows that the formal use of the 
exact Coloumb integral $\C'_g$, given by Eq.(\ref{ColoumbGa}), 
of the {\it Gaussian} screened Coloumb potential, 
instead of $\C'_e$, in Eq.(\ref{EmolE}) gives us 
approximately the same fits. Namely, the best fit of the energy 
is achieved at $\lambda=1/5$, with $r_c=0.1042$, optimal $R=1.323$, 
and minimal $E_{mol}=-1.172$. Also, the best fit of $R =1.40$ is
at $\lambda=1/29$, for which $r_c=0.0147$ and minimal $E_{mol}=-1.144$.
Here, we have used the same exchange integral as it is for the case
of exponential screened potential so these fits have been presented just 
for a comparison with our basic fits, and to check the results.
Note that for the case of Gaussian screened Coloumb integral 
we have no restriction on the allowed values of $\lambda$. Analysis
shows that, at big values of $\lambda$, e.g. at $\lambda>4$,
the integral $\C'_g$, given by (\ref{ColoumbGa}), 
rapidly oscillates in the region of small $\rho$ ($\rho<0.5$).
This means that when the correlation length $r_c$ becomes comparable
to the internuclear distance an effect of instability of the molecule
arises. This can be viewed as a natural criterium to fix the
upper limit of $\lambda$. Normally, we use the values $\lambda<1$,
for which case there are no any oscillations of $\C'_g$ (see Fig.~9).

{\small
\begin{table}[thp]\l{TableEnergy}
\begin{center}
\begin{tabular}{|c|c|c|c|}
\hline
$\lambda^{-1}$& $r_c$, a.u. & $R_{opt}$, a.u. & $E_{min}$, a.u. \\
\hline
 4 & 0.10337035071618050 & 1.297162129235449 & -1.181516949656805 \\ 
 5 & 0.08329699109108888 & 1.318393698326879 & -1.172984902150024 \\ 
 6 & 0.06975270534273319 & 1.333205576478603 & -1.167271240301846 \\ 
 7 & 0.05999677404817234 & 1.344092354783681 & -1.163188554065554 \\ 
 8 & 0.05263465942162049 & 1.352417789644028 & -1.160130284706318 \\ 
 9 & 0.04688158804756491 & 1.358984317233049 & -1.157755960428922 \\ 
10 & 0.04226204990365446 & 1.364292909163710 & -1.155860292450436 \\ 
11 & 0.03847110142927672 & 1.368671725082009 & -1.154312372623724 \\ 
12 & 0.03530417706681329 & 1.372344384866235 & -1.153024886026671 \\ 
13 & 0.03261892720535206 & 1.375468373051375 & -1.151937408039373 \\ 
14 & 0.03031323689615631 & 1.378157728092548 & -1.151006817317425 \\ 
15 & 0.02831194904031777 & 1.380497017045902 & -1.150201529091051 \\ 
16 & 0.02655851947236431 & 1.382550255552670 & -1.149497886394651 \\ 
17 & 0.02500959113834722 & 1.384366780045693 & -1.148877823925501 \\ 
18 & 0.02363136168905809 & 1.385985219224291 & -1.148327310762828 \\ 
19 & 0.02239708901865092 & 1.387436244558651 & -1.147835285349041 \\ 
20 & 0.02128533948435381 & 1.388744515712491 & -1.147392910500336 \\ 
21 & 0.02027873303335994 & 1.389930082626193 & -1.146993041730378 \\ 
22 & 0.01936302821907175 & 1.391009413196452 & -1.146629840949675 \\ 
23 & 0.01852644434336641 & 1.391996158084790 & -1.146298491232105 \\ 
24 & 0.01775915199935013 & 1.392901727808297 & -1.145994983116511 \\ 
25 & 0.01705288514774330 & 1.393735733699196 & -1.145715952370148 \\ 
26 & 0.01640064219648127 & 1.394506328745493 & -1.145458555325045 \\ 
27 & 0.01579645313764336 & 1.395220473843219 & -1.145220372020229 \\ 
28 & 0.01523519631632570 & 1.395884147817973 & -1.144999330178493 \\ 
29 & 0.01471245291356761 & 1.396502514589167 & -1.144793644973560 \\ 
30 & 0.01422439038752817 & 1.397080057337240 & -1.144601770891686 \\ 
\hline
\end{tabular}
\caption{The total minimal energy $E_{min}$ and the optimal internuclear 
distance $R_{opt}$ as functions of the correlation length $r_c$. 
The exponential screened Coloumb potential $V_e$ case 
(see Figures~15 and 16).}
\end{center}
\end{table}}

{\small
\begin{table}[thp]
\begin{center}
\begin{tabular}{|c|c|c|c|}
\hline
$\lambda^{-1}$& $r_c$, a.u. & $R_{opt}$, a.u. & $E_{min}$, a.u. \\
\hline
31 & 0.01376766836566138 & 1.397620687025853 & -1.144422362947838 \\ 
32 & 0.01333936209977966 & 1.398127830817745 & -1.144254245203342 \\ 
33 & 0.01293689977547854 & 1.398604504597664 & -1.144096385030938 \\ 
34 & 0.01255801083612469 & 1.399053372836414 & -1.143947871939897 \\ 
35 & 0.01220068312791624 & 1.399476798299823 & -1.143807900045981 \\ 
36 & 0.01186312715793131 & 1.399876883556063 & -1.143675753475045 \\ 
37 & 0.01154374612489787 & 1.400255505817128 & -1.143550794143290 \\ 
39 & 0.01095393745919852 & 1.400954915288619 & -1.143320213707519 \\ 
40 & 0.01068107105944273 & 1.401278573036792 & -1.143213620508321 \\ 
41 & 0.01042146833640030 & 1.401586548200467 & -1.143112256673494 \\ 
42 & 0.01017418516195214 & 1.401879953246168 & -1.143015746732479 \\ 
43 & 0.00993836493541500 & 1.402159797887369 & -1.142923750307661 \\ 
44 & 0.00971322867044429 & 1.402427000676349 & -1.142835958109381 \\ 
45 & 0.00949806639934841 & 1.402682399061957 & -1.142752088467028 \\ 
46 & 0.00929222969498477 & 1.402926758144872 & -1.142671884314343 \\ 
47 & 0.00909512514431396 & 1.403160778323019 & -1.142595110561057 \\ 
48 & 0.00890620863525624 & 1.403385101987775 & -1.142521551794315 \\ 
49 & 0.00872498034101540 & 1.403600319405678 & -1.142451010262626 \\ 
50 & 0.00855098030451296 & 1.403806973898863 & -1.142383304102633 \\ 
51 & 0.00838378454080327 & 1.404005566419838 & -1.142318265775268 \\ 
52 & 0.00822300158793934 & 1.404196559601683 & -1.142255740683024 \\ 
53 & 0.00806826944722482 & 1.404380381352424 & -1.142195585944305 \\ 
54 & 0.00791925286251402 & 1.404557428052374 & -1.142137669304475 \\ 
55 & 0.00777564089552400 & 1.404728067404676 & -1.142081868166104 \\ 
56 & 0.00763714476025456 & 1.404892640982100 & -1.142028068723488 \\ 
57 & 0.00750349588477794 & 1.405051466507240 & -1.141976165188595 \\ 
58 & 0.00737444417302681 & 1.405204839898059 & -1.141926059097351 \\ 
59 & 0.00724975644291090 & 1.405353037106507 & -1.141877658686723 \\ 
60 & 0.00712921502024112 & 1.405496315774223 & -1.141830878334298 \\ 
\hline
\end{tabular}
\caption{The total minimal energy $E_{min}$ and the optimal internuclear 
distance $R_{opt}$ as functions of the correlation length $r_c$. 
The exponential screened Coloumb potential $V_e$ case 
(see Figures~15 and 16).}
\end{center}
\end{table}}
\clearpage
%
\begin{figure}
\plot{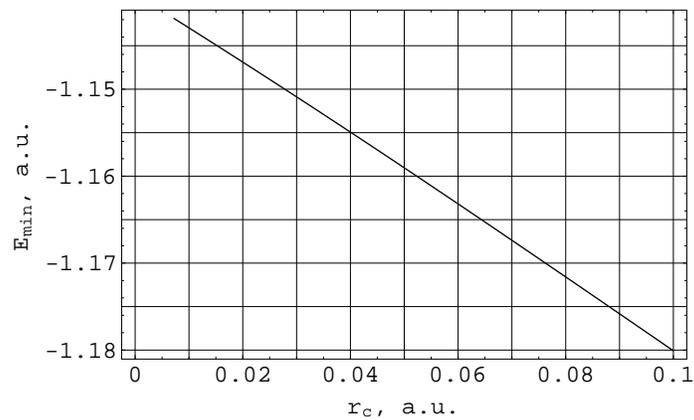}
\caption{The total minimal energy $E_{min}$ as a function
of the correlation length $r_c$. 
The exponential screened Coloumb potential $V_e$ case (see Tables~2 and 3).}
\label{Fig15}
\end{figure}
%
\begin{figure}
\plot{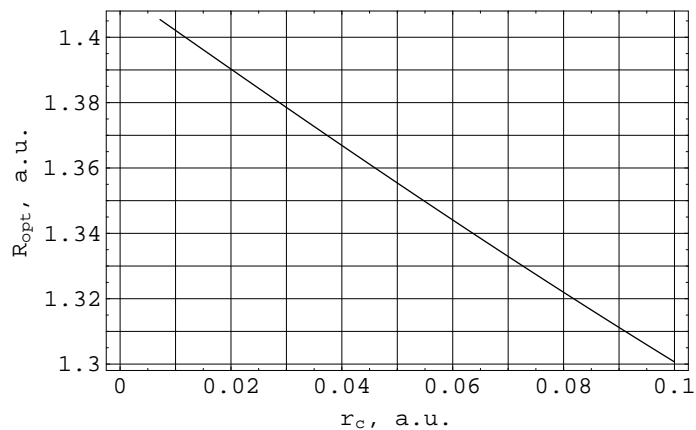}
\caption{The optimal internuclear distance $R_{opt}$  as a function
of the correlation length $r_c$. 
The exponential screened Coloumb potential $V_e$ case (see Tables~2 and 3).}
\label{Fig16}
\end{figure}

\end{document}